\documentclass[10pt]{article}

\usepackage{cite} 
\usepackage{amsmath}
\usepackage{hyperref}
\usepackage{cite}
\usepackage{amsmath,amssymb}
\usepackage{xcolor}
\usepackage{authblk}
\usepackage{graphicx}
\usepackage[shortlabels]{enumitem}
\usepackage{multirow}
\usepackage{floatrow}
\usepackage{subcaption}

\def\be{\begin{equation}} 
\def\ee{\end{equation}}   
\def \eea{\end{eqnarray}}
\def \bea{\begin{eqnarray}}

\newcommand{\dx}{ \dot{x} }

\newcommand{\Wr}{ \Omega_r }
\newcommand{\Wm}{ \Omega_m }
\newcommand{\Wl}{ \Omega_\Lambda }

\newcommand{\LCDM}{\text{LCDM}}

\definecolor{armygreen}{rgb}{0.29, 0.33, 0.13}

\newcommand{\uantof}{Departamento de Física, Universidad de Antofagasta, Aptdo. 02800, Chile}

\newcommand{\puc}{Instituto de F{\'i}sica, Pontificia Cat{\'o}lica Universidad de Chile, Av. Vicu{\~n}a Mackenna 4860, Santiago, Chile}
\newcommand{\viena}{Institut f\"ur Theoretische Physik,
 Technische Universit\"at Wien,
 Wiedner Hauptstrasse 8-10,
 A-1040 Vienna, Austria  and
 Atominstitut, Stadionalle 2,
 A-1020 Vienna, Austria}
 
\newcommand{\radboud}{Institute for Mathematics, Astrophysics and Particle Physics (IMAPP), Radboud University, Heyendaalseweg 135, 6525 AJ Nijmegen,The Netherlands}

\newcommand{\pucv}{Sede Esmeralda, Universidad de Tarapac\'a. Avda. Luis Emilio Recabarren 2477, Iquique, Chile.}

\begin{document}
\title{Statefinder analysis of scale-dependent cosmology}
\author[1]{Pedro D. Alvarez \thanks{E-mail: \href{mailto:pedro.alvarez@uantof.cl}{\nolinkurl{pedro.alvarez@uantof.cl}}}}
\affil[1]{\uantof}

\author[2,3]{Benjamin Koch \thanks{E-mail: \href{mailto:benjamin.koch@tuwien.ac.at}{\nolinkurl{benjamin.koch@tuwien.ac.at}}}}
\affil[2]{\puc}
\affil[3]{\viena}

\author[4]{Cristobal Laporte \thanks{E-mail: \href{mailto:calaporte@uc.cl}{\nolinkurl{cristobal.laportemunoz@ru.nl}}}}
\affil[4]{\radboud}

\author[2]{Felipe Canales \thanks{E-mail: \href{mailto:calaporte@uc.cl}{\nolinkurl{facanales@uc.cl}}}}

\author[5]{\'Angel Rinc\'on \thanks{E-mail: \href{mailto:aerinconr@academicos.uta.cl}{\nolinkurl{aerinconr@academicos.uta.cl}}}}
\affil[5]{\pucv}

\date{Received: date / Revised version: date}

\maketitle

\begin{abstract}
We study the statefinder parameters of a cosmological model based on scale-dependent gravity. The effective Einstein field equations come from an average effective action. From the dynamical system, we derive analytical expressions that improve the convergence of the numerical solutions. We determine the statefinder parameters for moderate redshift and compare them with well-known alternatives to $\Lambda$CDM.


\end{abstract}

\tableofcontents

\section{Introduction}\label{Intro}


Many predictions of Einstein's General Relativity \cite{GR} (GR) have been confirmed over the last 100 years.
The most prominent of these tests are the classical tests and solar system tests \cite{tests1,tests2,tests3},
direct detections of gravitational waves \cite{ligo1,ligo2,ligo3,ligo4,ligo5}, from black hole (BH) and neutron star mergers,
and the first image  of a black hole \cite{L1,L2,L3,L4,L5,L6}.



Shortly after the discovery of 
this successful theory as best theoretical description of gravitational phenomena, the question arose, whether this theory could fit into the framework of quantum field theory (QFT). Many promising attempts in this direction have been made. However, the quest is still open. This is in part due to the complexity of the theoretical framework and in part due to the lack of testable predictions and conclusive observational evidence.

Effective field theories (EFT) have proven to be an ideal theoretical instrument that allows to derive quantitative and testable predictions. This is both true in QFT and GR. Prominent approaches in GR include perturbative quantum gravity~\cite{Donoghue:1994dn} and the non-perturbative asymptotic safety paradigm~(for reviews see \cite{Bern:2002kj} and \cite{Niedermaier:2006wt,Eichhorn:2018yfc}).  A common feature of these EFT approaches is that they all have effective actions with scale-dependent (SD) couplings.
Thus, studying theories with SD is an
efficient way of exploring the phenomenological consequences of quantum field theories. 
This is also true for quantum gravity, where a full predictive quantum description is not yet available, but still exploring SD effects can give valuable phenomenological insight.
In this context, after the necessary scale setting,
working with the SD
gravitational couplings $G(x)$ (Newtons' coupling) and $\Lambda(x)$ (cosmological coupling), in the Einstein-Hilbert truncation is particularly useful. 
In this truncation, one 
incorporates aspects of quantum gravity (through SD) while still keeping the field equations second order. These second-order field equations can be complemented by an additional energy condition in such a way that they form a closed system that maintains general covariance \cite{Koch:2014joa,Rincon:2017ayr}.

With this and related frameworks it was possible to obtain numerous novel results for SD black holes \cite{Contreras:2013hua,Koch:2014cqa,Koch:2015nva,Koch:2016uso,Rincon:2017ypd,Rincon:2017goj,Rincon:2017ayr,Contreras:2017eza,Contreras:2018dhs,Contreras:2018gpl,Contreras:2018gct,Rincon:2018lyd,Rincon:2018dsq,Contreras:2018swc,Rincon:2018sgd,Rincon:2019cix,Contreras:2019cmf,Contreras:2019fwu,Bargueno:2020wlz,Rincon:2021hjj,Borissova:2022jqj}, and effects of SD in cosmology \cite{Bonanno:2001xi,Koch:2010nn,Saltas:2015vsc,Canales:2018tbn,Bertini:2019xws,Platania:2019qvo,Bargueno:2021nuc,Platania:2020lqb}. In a recent work we have developed a framework to study observational aspects of this approach in late-time cosmology by addressing the Hubble tension in terms of SD \cite{Alvarez:2020xmk}.


This pleasing progress with SD cosmology is to be seen in context of numerous other models that also give viable phenomenology of the late evolution of the Universe \cite{Peracaula:2022vpx,Abdalla:2022yfr,Moreno-Pulido:2022phq,Singh:2021jrp,Rezaei:2021qwd,SolaPeracaula:2021gxi,Panotopoulos:2020kpo,Panotopoulos:2018sso}.
Naturally, it would be interesting to investigate the similarities and differences of the SD approach with each of these models. 
However, this would mean that one would have to compare observable by observable across all these models.
This is an increasingly costly task since the number of necessary comparisons grows factorial with the number of models and observables. This is where model-independent parametrizations of the cosmological evolution can provide a helpful tool to simplify this task. 
In particular, the statefinder diagnostic gives a common ground for the comparison of multiple cosmological models \cite{Sahni:2002fz,Alam:2003sc,Zhang:2005yz,Evans:2004iq,Li:2014mua}.

Further, the statefinder diagnostic is useful for discriminating cosmological models. The reason is that even models with very similar expansion histories can have
discrepancies at higher order, which are perceivable in the statefinder approach. 
Thus, as statefinder diagnostic accounts for high order correction in derivatives on the scale factor, the approach is well suited to discriminate between seemingly equivalent models.

\subsection{Idea and structure of this paper}

The main idea of the paper is the following. Firstly, we reformulate the SD approach to cosmology to be suitable for a large-scale numerical study. The next step is to perform numerical analysis and compare it with related models. Finally, we analyze the results of this study by using the statefinder formalism, comparing our results with another coming from well-known papers.

The paper is organized as follows: in section 2, we present a brief self-contained discussion of the statefinder diagnostic. Then, in section 3, we summarize the statefinder approach in scale-dependent gravity, including the corresponding differential equations describing SD cosmology, and derive helpful formulas that improve the convergence of numerical solutions. Also, we compute the statefinder parameters as functions of redshift and compare them to well-known alternative cosmological models.
Finally, we summarize with the take-home message and final remarks in section 4.


\section{The motivation behind statefinder diagnostic}

A large variety of cosmological models can be used to derive the expansion history of the Universe. Since, in many cases, the underlying motivation for each model is quite diverse, it becomes useful to have a purely geometrical way of characterizing models. Let us consider a Taylor expansion of the scale factor (the only free function in the FRW metric),
\begin{equation}
    a(t)=a(t_0)+\dot{a}(t_0)(t-t_0)+\frac{1}{2!}\ddot{a}(t_0)(t-t_0)^2+\frac{1}{3!}\dddot{a}(t_0)(t-t_0)^3+\cdots
\end{equation}
The first derivative is related to the most important observable in cosmology, the Hubble parameter $H(z)$, and with the second derivative, we can construct a dimensionless cosmological function, the deceleration parameter,
\begin{equation}
 q \equiv - \frac{\ddot{a}/a}{H^2}\,, 
\end{equation} 
$q<0$ describes an accelerating universe. The deceleration parameter can also be expressed in terms of derivatives of $H$,
\begin{equation}
    q(z)=-1 +\frac{H'(z)}{H(z)}(1+z) \,,
\end{equation}
and therefore, $H(z)$ and $q(z)$ are observable because they can be inferred by luminosity distance or other measurements.
Within the context of the $\Lambda$CDM there is a large degeneracy in the deceleration parameter that tells us that measuring $q_0$ is not enough to discriminate $\Lambda$CDM models with different values of the parameters $(\Omega_m,\Omega_\Lambda)$.
Thus, as different models could coincide in both $H(z)$ and $q(z)$, further geometrical parameters, the statefinder parameters, were proposed \cite{Sahni:2002fz,Alam:2003sc}.
%
%
The statefinder parameters, $r$ and $s$, are  defined as follows
\begin{eqnarray}
r & = & \frac{\dddot{a}}{a H^3}\,, \\
s & = & \frac{r-1}{3 (q-\frac{1}{2})}\,, \label{exps}
\end{eqnarray}
where the dot represent differentiation with respect to the cosmic time $t$, $H=\dot{a}/a$ is the Hubble parameter, and $q=- \ddot{a}/(a H^2)$ is the decelerating parameter. 
By combining $q(z)$ with $r(z)$ or $s(z)$ we have a diagnostic pair that is very good at discriminating cosmological models. For example, from the Friedmann equation, we can see that the parameter $s$ is extremely sensitive to the total pressure
\begin{equation}\label{spcomment}
    s=\frac{\rho+p}{p}\frac{\dot{p}}{\dot{\rho}}\,.
\end{equation}
Therefore $s \rightarrow \infty$ when $p \rightarrow 0$. As pointed out in \cite{Alam:2003sc}, for $\Lambda$CDM this occurs for $z_\ast \sim 10$. Therefore the parameter $s$ diagnoses the presence of dark energy even at higher redshifts when the contribution of dark energy to the total energy is small. Also, the definition of the statefinder parameter $r$ is motivated by removing the dependence on dark energy.

Different trajectories of the $q-r$ plane or the $q-s$ plane correspond to different dark matter and dark energy models. Thus, studying this ``phase space'' helps to identify the main features of a given cosmological model without any reference to the microscopic degrees of freedom. Notice that the pair $\{q,r\}$ or $\{q,s\}$ provide a good way of establishing the ``distance" between two cosmological models.
Let us define "the classical limit" of the SD cosmological model as the case when $\dot{g}(t_0)=0$ (this definition is justified below). Such a case corresponds to the pair $\{r, s\} \equiv \{1,0\}$, i.e., it corresponds to a spatially flat $\Lambda$CDM Universe up to a possible renormalization of the absolute value of the Newton constant ($g(t_0)$ may or may not be equal to 1).
%

In order to contrast with observations, there are two main strategies in the literature. Firstly, for concrete models, $q(z)$, $r(z)$, and $s(z)$ can be inferred from the best fit values of the parameters of the model that are implied from observations. 
Such analysis have been applied to several dark energy models \cite{Kumar:2011sw, Giostri:2012ek,Cardenas:2013roa,Rani:2014sia, Camarena:2019moy}, see also \cite{AlMamon:2018uby} and references therein. Secondly, there are also attempts to reconstruct the parameters as functions of redshift geometrically. Although this is a hard task, there has been recent progress in this direction \cite{Zhai:2013fxa,AlMamon:2018uby,Mukherjee:2021ggf}. This work will take advantage of this formalism to answer if the scale-dependent gravity of \cite{Alvarez:2020xmk} may correspond to a dynamically equivalent cosmological model. In this paper, we provide an answer to this question: scale-dependent gravity is an independent model, as can be seen by looking at the state finder parameters derived below.

\section{Statefinder parameters in the scale-dependent cosmological scenario} \label{sfpmts}

The system of differential equations that determines the cosmological background evolution is formed by three equations \cite{Alvarez:2020xmk}
\begin{align}
&\frac{1}{H_0^2}\left(H^2 - H \frac{\dot g}{g}\right)=\Wl \lambda(t) + \frac{\Wr}{a^{4}} g + \frac{\Wm}{a^{3}} g \,,\label{SD1}\\
&\frac{1}{H_0^2}\left(2  \dot H + 3H^2-2H\frac{\dot g}{g}+2\frac{\dot g^2}{g^2} - \frac{\ddot g}{g} \right)=3 \Wl \lambda(t) - \frac{\Wr}{a^{4}} g\,,\label{SD2}\\ 
& \frac{\ddot g}{g} -H\frac{\dot{g}}{g} - 2 \frac{\dot g^2}{g^2}=0\,,\label{NECCosm}
\end{align}
where
\begin{equation}\label{dimless}
 g(t)=\frac{G(t)}{G_0}\,, \quad \lambda(t)=\frac{\Lambda(t)}{\Lambda_0}\,.
\end{equation}
As in the $\Lambda$CDM model, the density parameters $\Wm$, $\Wr$, and $\Wl$ describe the contents of the Universe. The time dependence of the $\Lambda$ coupling that is implied in the scale--dependent scenario results in time-dependent dark energy
\begin{equation}
 \Wl \lambda(t)=\frac{\Lambda(t)}{3H_0^2}\,.
\end{equation}

We will explore the phase space of the dynamical system given by Eqs. (\ref{SD1}-\ref{NECCosm}). For this discussion, it is convenient to use the dimensionless functions (\ref{dimless}) and
\begin{equation}
 x=\frac{a(t)}{a_0}\,.
\end{equation}
The scale-dependent cosmological equations take the form
\begin{align}
 &\frac{x'(\tau)^2}{x(\tau)^2}-\frac{g'(\tau) x'(\tau)}{g(\tau) x(\tau)}= \Wm g(\tau)x(\tau)^{-3}+\Wr g(\tau)x(\tau)^{-4}+\Wl \lambda(\tau)\,,\label{eq1}\\
 &\frac{x'(\tau)^2}{x(\tau)^2}-2\frac{g'(\tau) x'(\tau)}{g(\tau) x(\tau)}+2\frac{g'(\tau)^2}{g(\tau)^2}+2\frac{x''(\tau)}{x(\tau)}-\frac{g''(\tau)}{g(\tau)}=-\Wr g(\tau) x(\tau)^{-4}+3\Wl \lambda(\tau)\,,\label{eq2}\\
 &\frac{g''(\tau)}{g(\tau)}-\frac{g'(\tau) x'(\tau)}{g(\tau) x(\tau)}-2\frac{g'(\tau)^2}{g(\tau)^2}=0\,.\label{eq3}
\end{align}
The evolution of this dynamical system can be determined by giving the initial conditions $x(\tau_0)$, $x'(\tau_0)$, $g(\tau_0)$ and $g'(\tau_0)$, where we denoted $\tau_0$ as the present value of the evolution coordinate $\tau$. This means an increase from one to four d.o.f. with respect to the $\Lambda$CDM case, where only $x'(\tau_0)$ is required to provide initial values. In the $\Lambda$CDM case, this degree of freedom is quite often traded for the value of $h$.

When switching between the time variables $t$ and $\tau$ is useful to have at hand the relations,
\begin{align}
 \dx(t)/x(t) =& (H_{100} h)  x'(\tau)/x(\tau)\,,\\
 t =& (H_{100} h)^{-1}  \tau\,.\label{tau}
\end{align}
In particular,
\begin{align}
 H_0 =& (H_{100} h)  x'(\tau_0)/x(\tau_0)\,,\\
 t_\text{age} =& (H_{100} h)^{-1}  \tau_\text{age}\,. \label{tage}
\end{align}

%
%

In \cite{Alvarez:2020xmk}, we found that the model (\ref{eq1})-(\ref{eq3}) can release the tension on the Hubble constant inferred from the CMB and low redshift observations. A natural question arises: how to discriminate the dynamics derived from this model from the plethora of cosmological models available to date?

To answer this question, we will study the statefinder parameters of the scale-dependent scenario. We will focus on the effects of different values of the initial conditions while keeping the density parameters $\Omega_m$, $\Omega_r$, and $\Omega_\Lambda$ fixed. We will consider the fiducial values
\begin{equation}
    \Omega_m=0.7\,, \quad \Omega_r=0\,, \quad \Omega_\Lambda=1-\Omega_m\,.
\end{equation}

Exact solutions for the SD model are not known in general, and therefore we will consider numerical solutions,
\begin{equation}
    (z, X) = (z(\tau), X(\tau))\,,
\end{equation}
where $X$ stands for $a$, $H$, $q$, $r$ or $s$. Moreover, numerical stability becomes a problem in the evaluation of $q$, $r$ and $s$ and therefore it is convenient to obtain analytic expressions for $H$, $q$ and $r$ that lower the order in derivatives of expressions that are evaluated numerically. We will do this by using the dynamical system (\ref{eq1})-(\ref{eq3}). From the NEC, eq. (\ref{eq3}) we get
\begin{equation}\label{numsta1}
    \frac{g'}{g}=\frac{g'_0}{g_0}\frac{g}{g_0}\frac{x}{x_0}\,.
\end{equation}
Also, substituting $g''$ from (\ref{eq3}) and $\lambda$ from (\ref{eq1}) into (\ref{eq2}) we get
\begin{equation}\label{numsta2}
    \left(\frac{x'}{x}\right)'=f(x,g)\,, 
\end{equation}
where
\begin{equation}
    f(x,g)=-2\Omega_r \frac{g}{x^4}-\frac{3}{2}\Omega_m\frac{g}{x^3}\,.
\end{equation}

The evaluation of the r.h.s. of equations (\ref{numsta1}), (\ref{numsta2}) has a much better numerical stability than the evaluation of the l.h.s. We will use this procedure to study $H$, $q$, $r$, and $s$ for the SD model, and we will compare to the reference models $\Lambda$CDM, Brans-Dicke, quiessence, and kiessence.

For the Hubble parameter we have
\begin{equation}
    H=h H_{100}\frac{x'}{x}\,,
\end{equation}
where
\begin{equation}\label{numsta3}
   \frac{x'}{x}= \frac{x'_0}{x_0}+\int_{\tau_0}^\tau d\tilde{\tau} f(x(\tilde{\tau}),g(\tilde{\tau})) \,.
\end{equation}
Evaluation of the r.h.s. of (\ref{numsta3}) is more time consuming than evaluation of the l.h.s. but it offers better numerical stability. In the computation of $H$, numerical stability does not pose a serious problem in most cases, but for the computation of higher-order parameters, it is essential to use expressions with better numerical stability.

The parameters $q$, $r$, and $s$ are dimensionless, meaning that their formal expressions in terms of time derivatives or derivatives with respect to $\tau$ remain the same,
\begin{equation}
    q=-\frac{x''/x}{(x'/x)^2}\,,
\end{equation}
and so on for $r$ and $s$. Using (\ref{numsta2}) we can rewrite $q$ as
\begin{equation}\label{qstable}
    q=-1-\frac{f(x,g)}{(x'/x)^2}\,,
\end{equation}
where in the denominator of the second term, we can either use the l.h.s. or the r.h.s. of (\ref{numsta3}), since in this case, the usage of l.h.s. of (\ref{numsta3}) does not introduce numerical instabilities in such term.

For $r$, which is defined by,
\begin{equation}
    r=\frac{x'''/x}{(x'/x)^3}\,,
\end{equation}
we get
\begin{equation}\label{rstable}
    r=-q+\left(4\Omega_r \frac{g}{x^4}+\frac{3}{2}\Omega_m\frac{g}{x^3}\right)(x'/x)^{-2}+\left(\frac{g'_0}{g_0}\frac{g}{g_0}\frac{x}{x_0}\right)f(x,g)\,.
\end{equation}
For numerical stability, it is sufficient to use (\ref{qstable}) in the first term and the l.h.s. of (\ref{numsta3}) in the second term. Substituting $q$ from (\ref{qstable}) in (\ref{rstable}) the $r$ parameter can be written as
\begin{equation}
    r=1+2\Omega_r \frac{g}{x^4}(x'/x)^{-2}+\left(\frac{g'_0}{g_0}\frac{g}{g_0}\frac{x}{x_0}\right)f(x,g)\,.
\end{equation}
From here, we see that $r$ will remain pegged at 1 if $g'_0=0$ for the entire matter-dominated era, similar to what happens in $\Lambda$CDM. Compared to the $\Lambda$CDM we see that the term accompanying $\Omega_r$ gets an extra $g(\tau)$ factor,
\begin{equation}
    r_{\Lambda CDM}=1+2\Omega_r x^{-4} (H_0/H(t))^{-2}\,.
\end{equation}
The parameter $s$ can be evaluated by substituting (\ref{qstable}) and (\ref{rstable}) in (\ref{exps}).

Note that by using (\ref{numsta1}) in (\ref{numsta2}) we can obtain expressions free of derivatives of $x$ and $g$ for higher-order parameters. These techniques allowed us to evaluate the statefinder parameters for higher redshift without the need for Pade approximants, see for instance \cite{Capozziello:2017ddd}.

\subsection{Comparison with other models}

We will compare the SD model with referential models such as $\Lambda$CDM, quiessence, kinnesence, polynomial dark energy, and Brans-Dicke cosmological models. Before doing this, let us use (\ref{numsta1}) to answer a frequently asked question about the scale-dependent model: is the SD cosmological model equivalent to a Brans-Dicke theory? The answer is no. We can use (\ref{numsta1}) to obtain an integral expression for $g(\tau)$,
\begin{equation}
    g=g_0\left(1-\frac{g'_0}{x_0 g_0}\int_{\tau_0}^\tau d\tilde{\tau} x(\tilde{\tau})\right)^{-1}\,.
\end{equation}
Therefore the dynamics of $g(\tau)$ predicted by the SD model is essentially different from the one predicted by the Brans-Dicke theory (BD), where
\begin{equation}\label{GBD}
   \left. \frac{G}{G_0}\right|_{\text{BD}}= x^{-\frac{1}{\omega+1}}\,,
\end{equation}
and $\omega$ is a parameter present in the BD theory \cite{Hrycyna:2013hla}. In the SD model, an initial condition $g'_0=0$ grants $g=const$ for all $t$, but such behavior can only be obtained by an improper limit of the BD theory $\omega \rightarrow \infty$. There are, however, values of the initial conditions of the SD model that can reproduce almost exactly $H(z)$ and the $q(z)$ parameter in the BD theory and differences show up at higher order only, the behavior of the  (see plots in Figures~\ref{varyapcase},~\ref{varyapcase_gpneg} and~\ref{varygpcase} below).

%

We will also refer to other well-known cosmological models for comparison. We will include in our discussion the quiessence and kinessence models already discussed in \cite{Sahni:2002fz,Alam:2003sc}. Let us recall the equation of state
\begin{equation}
    p=w \rho\,.
\end{equation}
In $\Lambda$CDM we have $w=-1$, in quiessence models we have $w<-1/3$ and in kinessence we have a time-dependent $w$ or $w(z) \ne const$. A particular realization of kinessence is the quintessence model where an extra scalar field with a self-interaction potential minimally coupled to gravity \cite{Sahni:1999gb, Sahni:2002kh}. In such models, the statefinder pair takes the form
\begin{align}
    r =& 1+\frac{9}{2}\Omega_X w(1+w)-\frac{3}{2}\Omega_X \frac{\dot{w}}{H}\,,\\
    s =& 1+w-\frac{1}{3}\frac{\dot{w}}{wH}\,,
\end{align}
where the function $w(t)$ can be related to a mixture of physical ($\Omega_X$) and geometrical parameters (in this case $q(t)$),
\begin{equation}
    w(t)=\frac{2q(t)-1}{3\Omega_X}\,.
\end{equation}
We will also compare our results to a spatially flat dust-filled Universe in the Brans-Dicke theory \cite{Goswami:2017pcy}, where the Hubble parameter is given by
\begin{equation}
    H=\frac{H_0}{\sqrt{1+\frac{5\omega+6}{6(\omega+1)^2}}}\sqrt{\Omega_m (1+z)^{\frac{3\omega+4}{\omega+1}}+\Omega_\Lambda}\,.
\end{equation}
Constraints on the rate of variation of Newton constant, given in (\ref{GBD}), provide that the parameter $\omega$ is constrained to $\omega>40000$ \cite{Bertotti:2003rm,DeFelice:2005bx}. Parameter estimation based on luminosity distances and the Union 2.1 compilation \cite{Susuki2012} gives the best fit parameters $\Omega_m = 0.296$ and $\Omega_\Lambda = 0.704$.

Finally, we will also compare to a polynomial dark energy model, where the dark energy density is expressed as a truncated Taylor series polynomial in $(1+z)$ \cite{Saini:1999ba},
\begin{equation}
    H(z)=H_0 \sqrt{\Omega_m (1+z)^3+\rho_{DE}}\,, \quad \rho_{DE}=A_1+A_2(1+z)+A_3(1+z)^2\,.
\end{equation}
The expression for the distance modulus is given by
\begin{equation}
    \mu(z)=5\log_{10} D_L(z)-5\log_{10} H_0 +52.38\,,
\end{equation}
while luminosity distance reads,
\begin{equation}
    \frac{D_L}{1+z}=\int_0^z \frac{dz}{\sqrt{\Omega_m (1+z)^3+A_1+A_2(1+z)+A_3(1+z)^2}}\,.
\end{equation}
We will use observational constraints coming from the following observations:
\begin{itemize}
    \item The latest $H(z)$ data, consisting in 28+1 data points given in \cite{Rani:2014sia}. Such dataset is based in 28 points given in \cite{Farooq:2013hq} plus the value of $H_0$ estimated in \cite{Riess:2011yx}.
    \item The Union 2.1 $\mu(z)$ data compilation consists of data for 833 SNe, drawn from 19 datasets. Of these, 580 SNe pass usability cuts \cite{Susuki2012}.
\end{itemize}
These observations are sufficient to constrain reasonably well the following restricted versions of the polynomial DE model:
\begin{align}
    A_1 =& 1 - \Omega_\text{mat}\,, A_2=0\,, \quad \text{, `model A'}\,; \label{modelAeq}\\
    A_3 =& 1 - \Omega_\text{mat}-A_1\,, A_2=0\,, \quad \text{, `model B'}\,.\label{modelBeq}
\end{align}
We define $\chi^2$ as
\begin{equation}
    \chi^2 = \sum_{i=1}^n \left(\frac{y(z)-y^\text{obs}_i}{\sigma_i}\right)^2\,,
\end{equation}
where $y(z)$ represents model predictions for $H(z)$ or $\mu(z)$.  We determined the best fit parameters, and reduced chi-squared $\chi^2/\nu$, where $\nu$ is the number of degrees of freedom, $\nu(H(z))=26$ and $\nu(\mu(z))=577$ for both models, See table \ref{tablepolDEresuls} with the results.

\begin{table}[h!]
\centering
\begin{tabular}{c c  c c  c c } 
\multicolumn{2}{c}{} &  \multicolumn{2}{c}{$H(z)$} & \multicolumn{2}{c}{$\mu(z)$} \\ 
\multicolumn{2}{c}{} &  B.F.  & M.V. &  B.F. & M.V. \\ \cline{3-6}
\multicolumn{2}{c}{} &  $\chi^2/\nu \approx 0.682$ &  &  $\chi^2/\nu \approx 0.974$ & \\\hline
\multirow{3}{*}{\begin{minipage}{12mm}
`model A'
\end{minipage}}  & $\Omega_\text{mat}$   & 0.261 & $0.26\substack{+0.02 \\ -0.02}$ & 0.281 & $0.21\substack{+0.10 \\ -0.12}$ \\ \cline{3-6}
                      & $H_0$   & 80.118 & $76.16\substack{+7.86 \\ -8.92}$ & 70.350 & $60.90\substack{+13.25 \\ -9.53}$ \\ \cline{3-6}
                      & $A_3$   & -0.184 & $-0.11\substack{+0.21 \\ -0.13}$ & -0.013 & $0.31\substack{+0.52 \\ -0.42}$ \\  \hline
                      \\
\multicolumn{2}{c}{} &  $\chi^2/\nu \approx 0.682$ &  &  $\chi^2/\nu \approx 0.974$ & \\ \hline
\multirow{3}{*}{\begin{minipage}{12mm}
`model B'
\end{minipage}}  & $\Omega_\text{mat}$   & 0.320 & $0.32\substack{+0.06 \\ -0.06}$ & 0.285 & $0.31\substack{+0.20 \\ -0.17}$ \\ \cline{3-6}
                      & $H_0$   & 72.354 & $72.11\substack{+2.04 \\ -2.05}$ & 69.880 & $69.92\substack{+0.43 \\ -0.42}$ \\ \cline{3-6}
                      & $A_1$   & 0.906 & $0.90\substack{+0.13 \\ -0.14}$ & 0.729 & $0.75\substack{+0.18 \\ -0.16}$ \\  \hline
\end{tabular}
\caption{$H_0$ is given in units of $[\text{km}\ \text{s}^{-1} \text{Mpc}^{-1}]$. The reduced chi-squared values take the same value when rounded to the third decimal for both models.}
\label{tablepolDEresuls}
\end{table}

\begin{figure}[ht!]
    \centering
        \begin{subfigure}{.35\textwidth}
        \includegraphics[trim={0.2cm 0.2cm 0.2 0.2cm},clip,width=1.\linewidth]{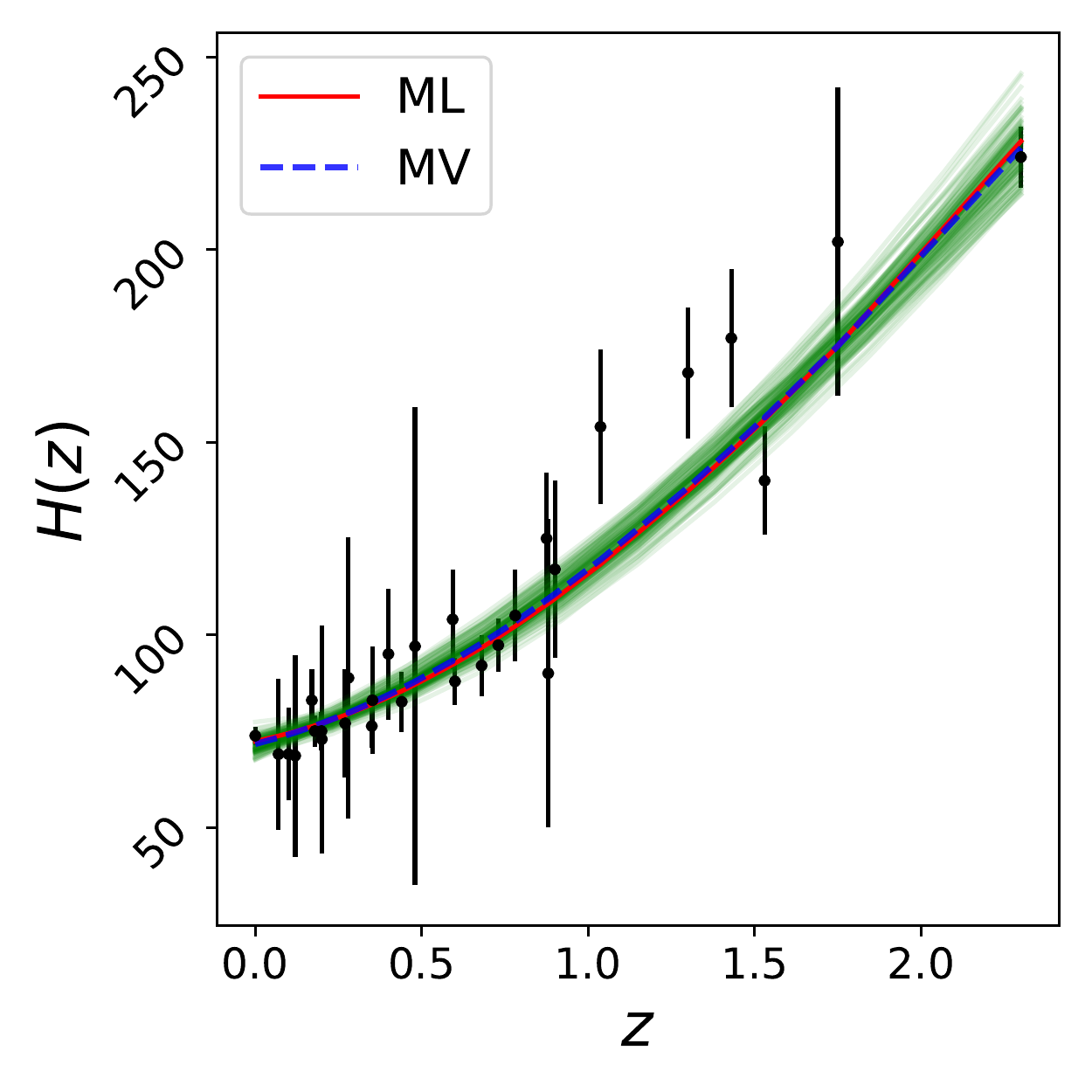}\\
        \includegraphics[trim={0.2cm 0.2cm 0.2 0.2cm},clip,width=1.\linewidth]{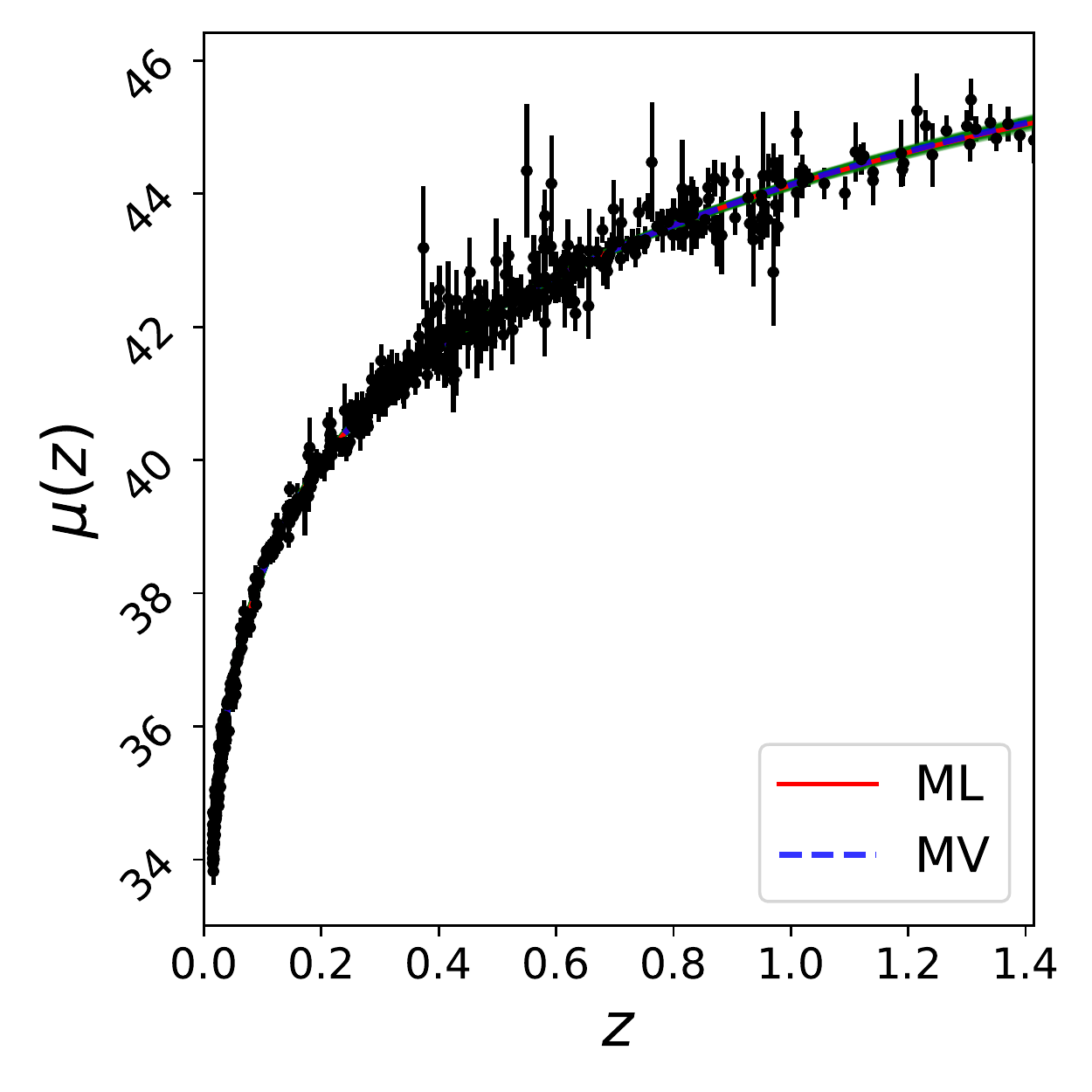} 
        \caption{Red is the BF value (ML) and blue-dashed the mean value (MV). In green we have a hundred randomly selected set of parameteres within 3$\sigma$ of the BF value of the posterior distribution.}
        \label{fig:H-mu-A}
        \end{subfigure}
        \begin{subfigure}{.55\textwidth}
        \includegraphics[trim={0.2cm 0.2cm 0.5 0.2cm},clip,width=1.\linewidth]{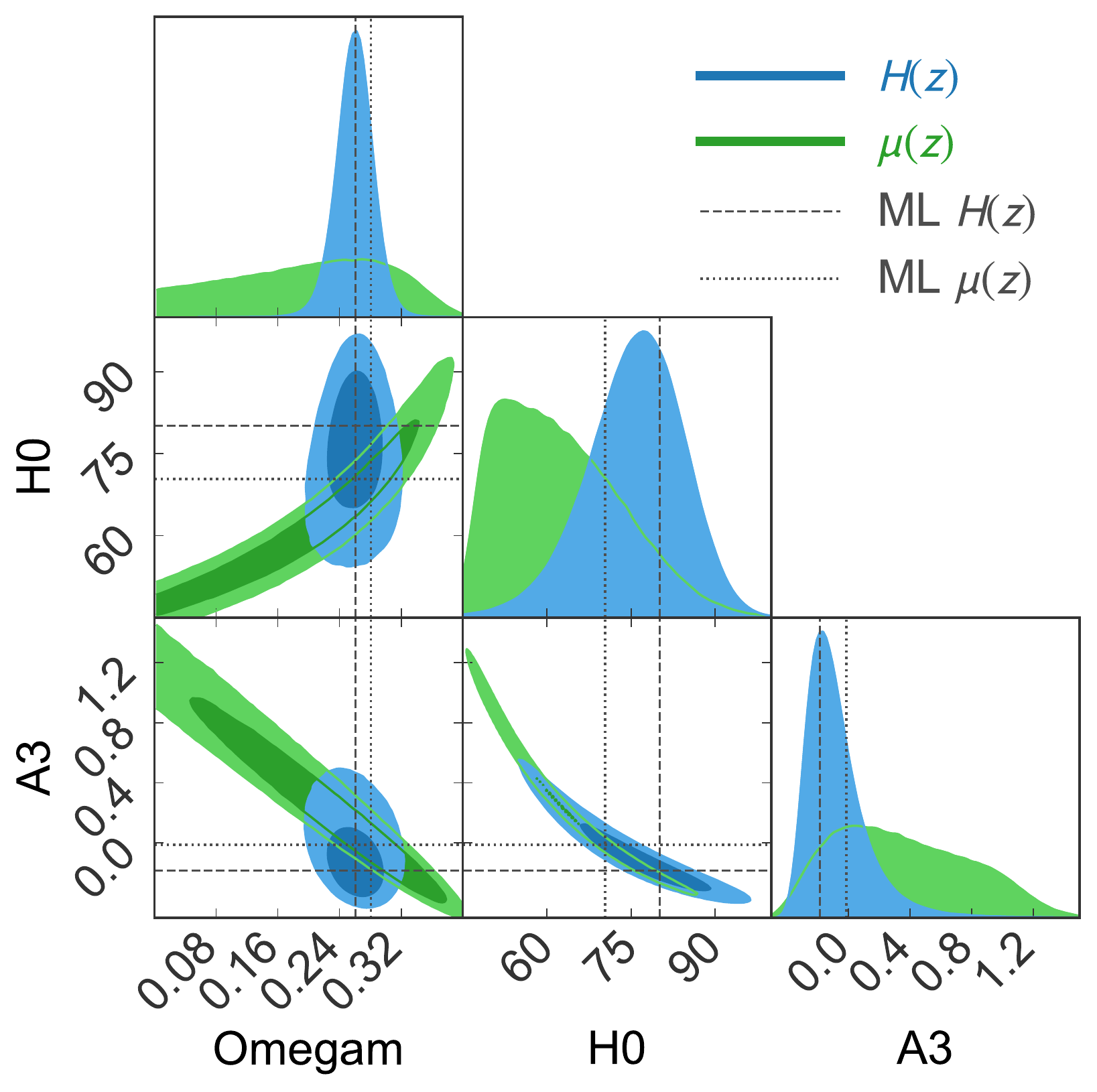}
        \caption{`Model A' is compatible with $H(z)$ data and $\mu(z)$ data. Marginalized posterior for $\mu(z)$ is non-Gaussian.}
        \label{corner-A}
        \includegraphics[width=1.\linewidth]{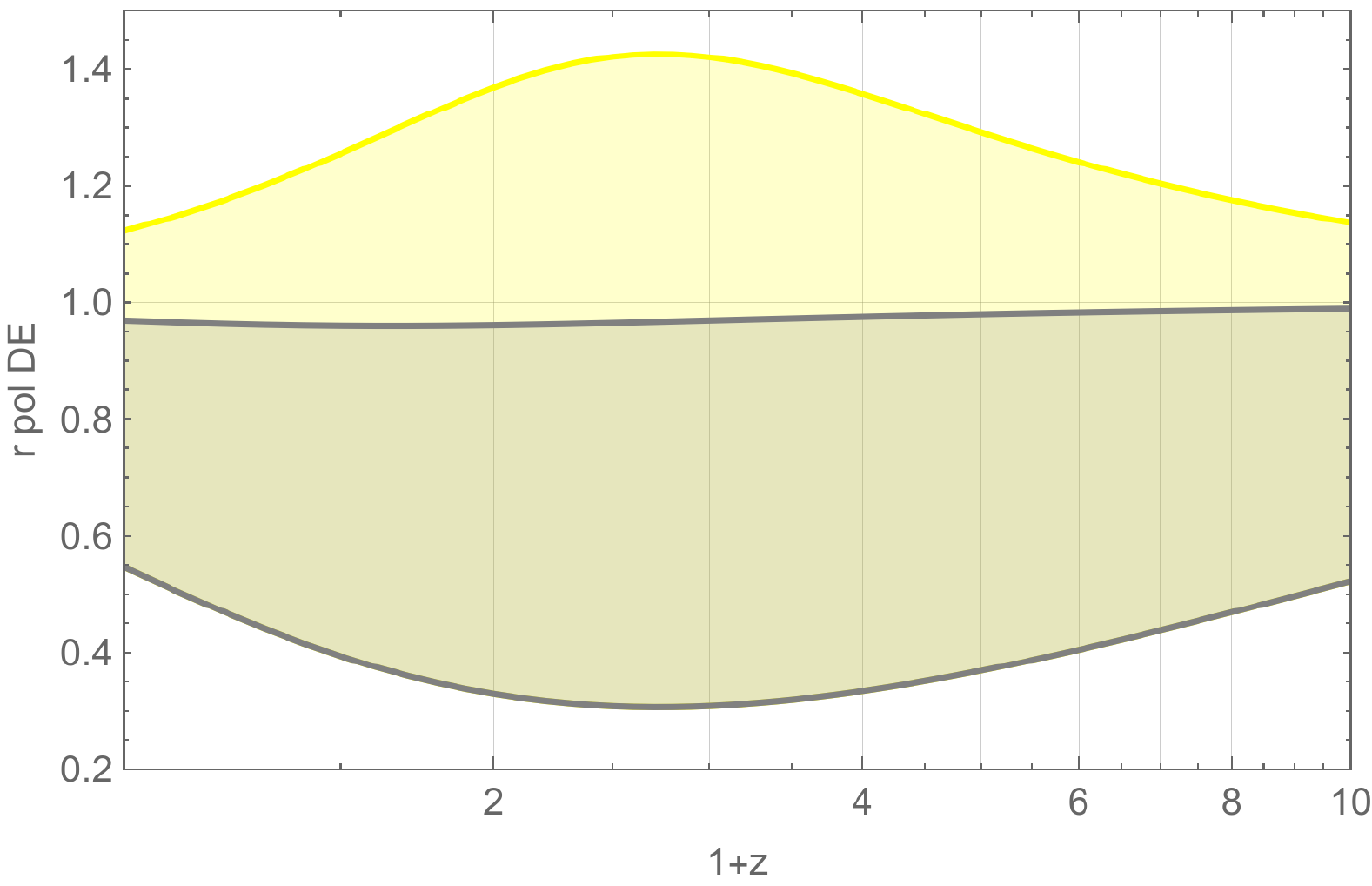}
        \caption{Yellow shaded region correspond to all models within 1$\sigma$ of the BF value for $\mu(z)$. The gray shaded region, a subset of the yellow shaded region, has the additional constraint of $s < 1$, see comments around eq. (\ref{spcomment}).}
        \label{r-A}
        \end{subfigure}\caption{`Model A' defined in eq. (\ref{modelAeq}). Best fit values are summarized in table \ref{tablepolDEresuls}.}
\label{modelAfig}
\end{figure}

\begin{figure}[ht!]
    \centering
        \begin{subfigure}{.35\textwidth}
        \includegraphics[trim={0.2cm 0.2cm 0.2 0.2cm},clip,width=1.\linewidth]{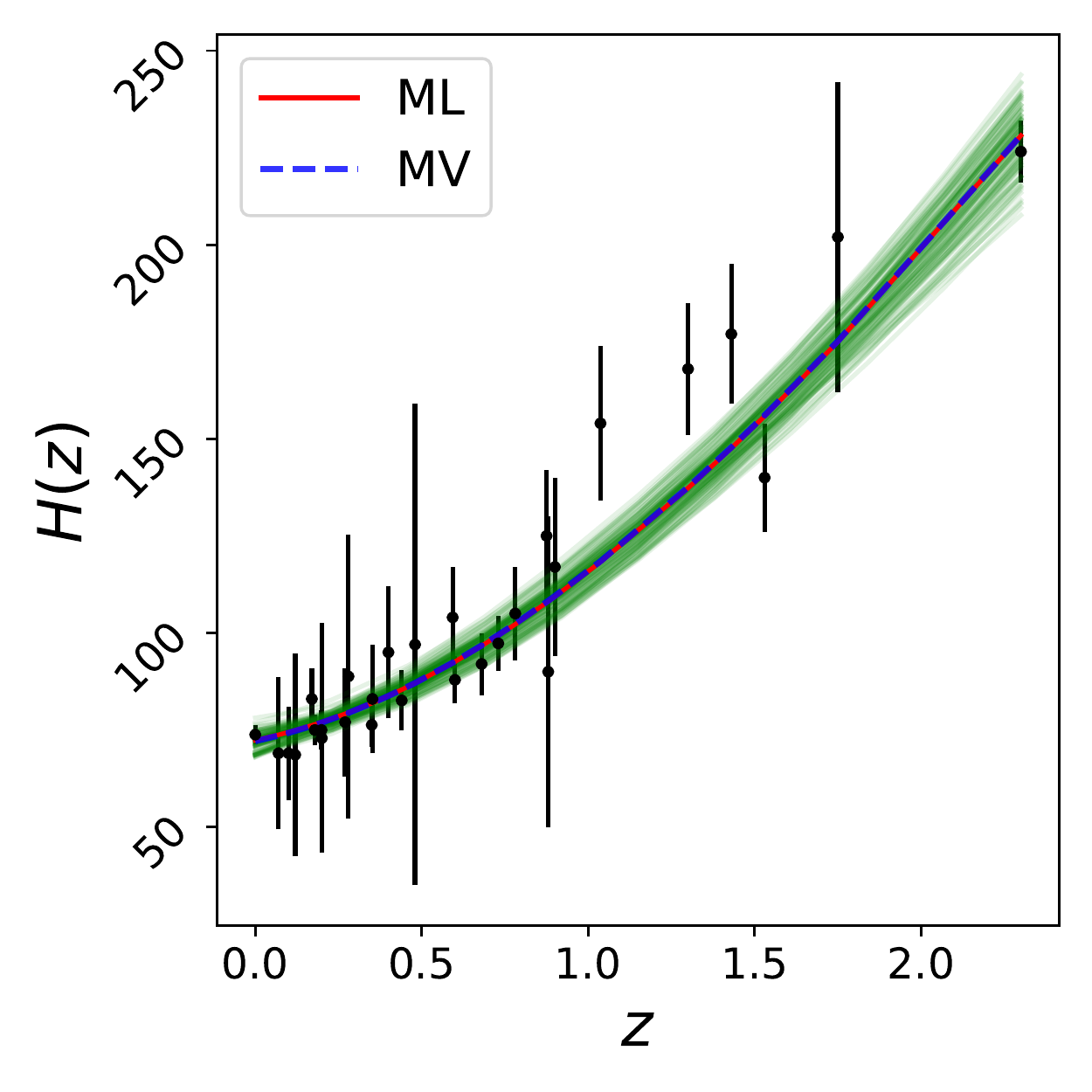}\\
        \includegraphics[trim={0.2cm 0.2cm 0.2 0.2cm},clip,width=1.\linewidth]{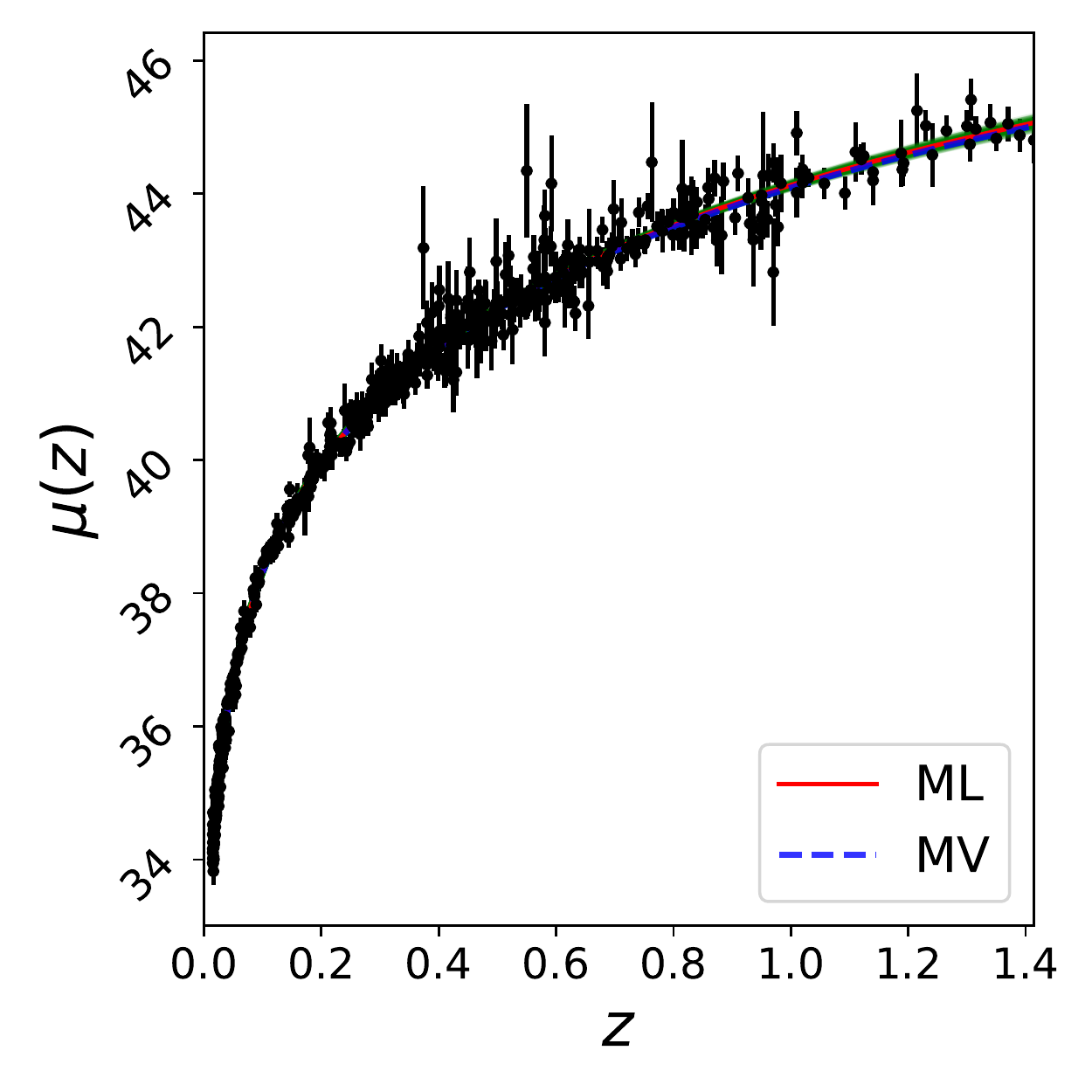} 
        \caption{Red is the BF value (ML) and blue-dashed the mean value (MV). In green we have a hundred randomly selected set of parameteres within 3$\sigma$ of the BF value of the posterior distribution.}
        \label{H-mu-B}
        \end{subfigure}
        \begin{subfigure}{.55\textwidth}
        \includegraphics[trim={0.2cm 0.2cm 0.5 0.2cm},clip,width=1.\linewidth]{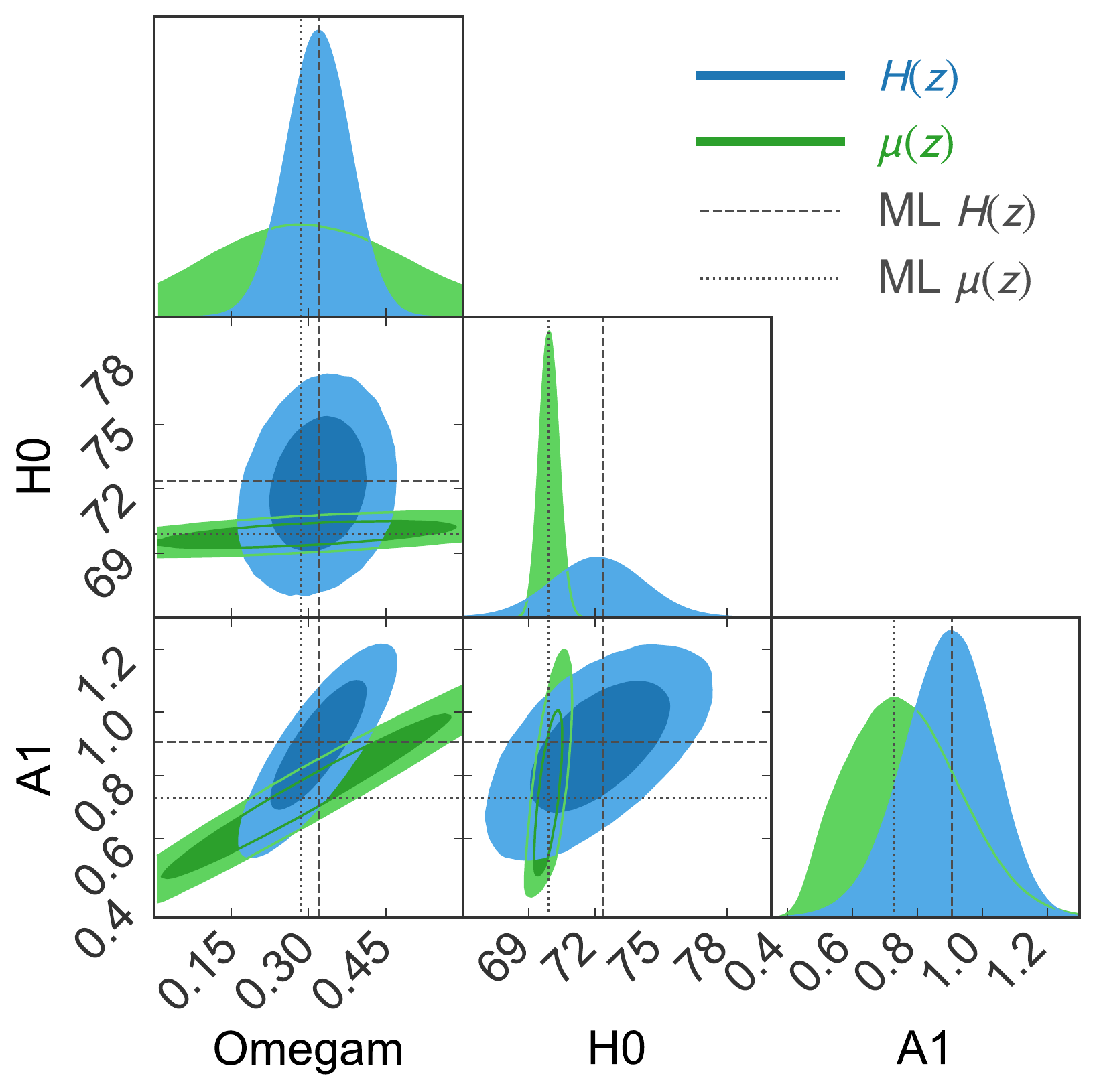}
        \caption{`Model B' is also compatible with $H(z)$ data and $\mu(z)$ data. Marginalized posterior for $\mu(z)$ is Gaussian.}
        \label{corner-B}
        \includegraphics[width=1.\linewidth]{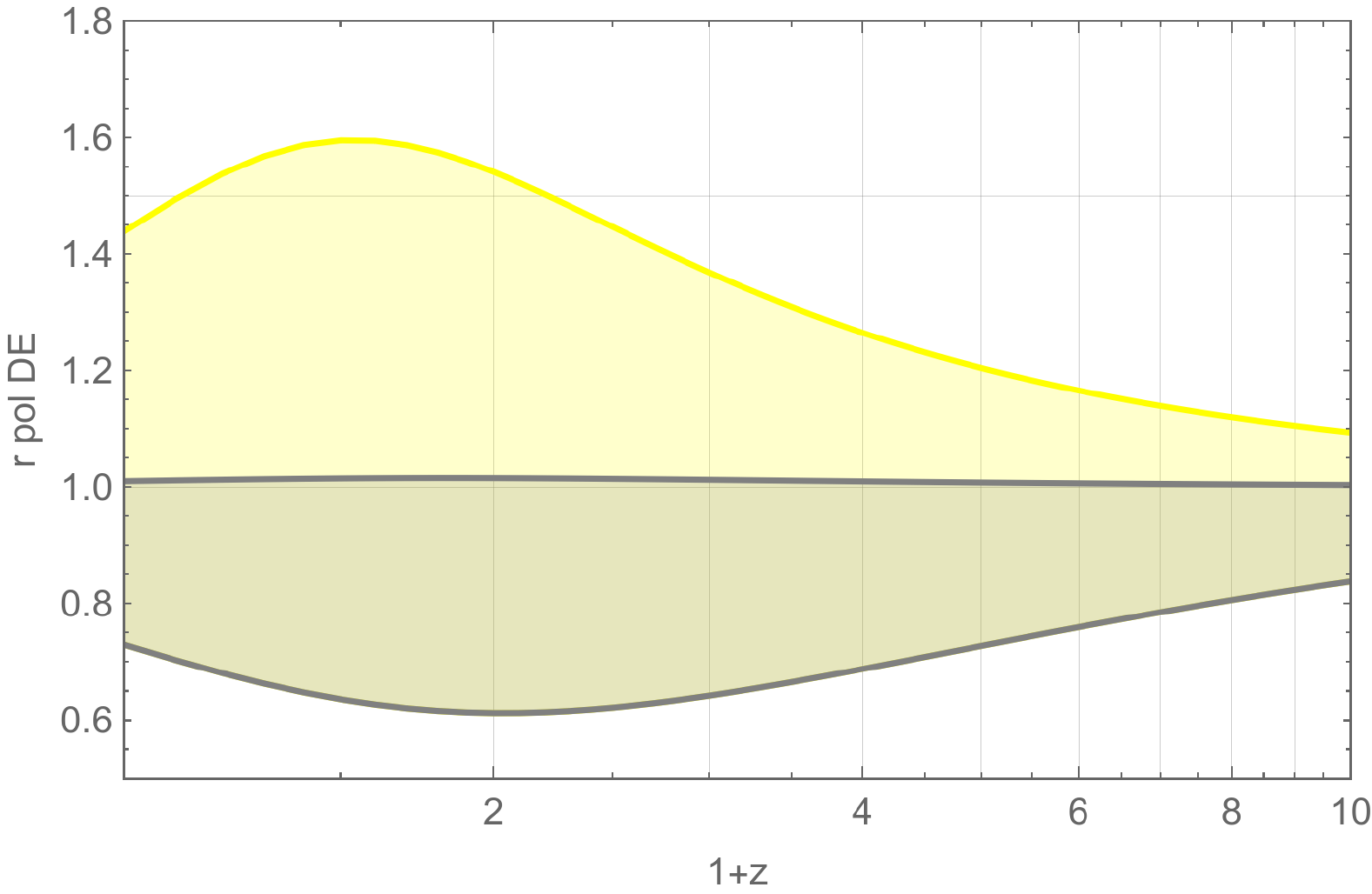}
        \caption{Yellow shaded region correspond to all models within 1$\sigma$ of the BF value for $\mu(z)$. The gray shaded region, a subset of the yellow shaded region, has the additional constraint of $s < 1$, see comments around eq. (\ref{spcomment}).}
        \label{r-B}
        \end{subfigure}
\caption{`Model B' defined in eq. (\ref{modelBeq}). Best fit values are summarized in table \ref{tablepolDEresuls}.}
\label{modelBfig}
\end{figure}

We used the EMCEE Python library to implement the Markov chain Monte Carlo ensemble sampler that allows computation of maximum likelihood contours (M.L.C.) \cite{Foreman-Mackey:2012any}. For 'model A', we obtain the results of figure \ref{modelAfig}, while for 'model B', the results are shown in figure \ref{modelBfig}. From the M.L.C. we can see that `model B' gets better constrained by the $H(z)$ and $\mu(z)$ data. Consequently, we will use the mean values within errors for `model B' to provide us a reference region.

In figures  \ref{r-A} and \ref{r-B} we have consider the mean values with 1$\sigma$ error inferred from the $\mu(z)$ data (yellow shaded region), see last column of table \ref{tablepolDEresuls}. The gray shaded region correspond to models such that $s(z) < 1$ when $z < 10$, which is a restriction that removes models that could alter the redshift value of the drag epoch.

\newpage
\subsection{Case $g'_0>0$}

\begin{figure}[ht]
  \centering
  \includegraphics[width=.49\linewidth]{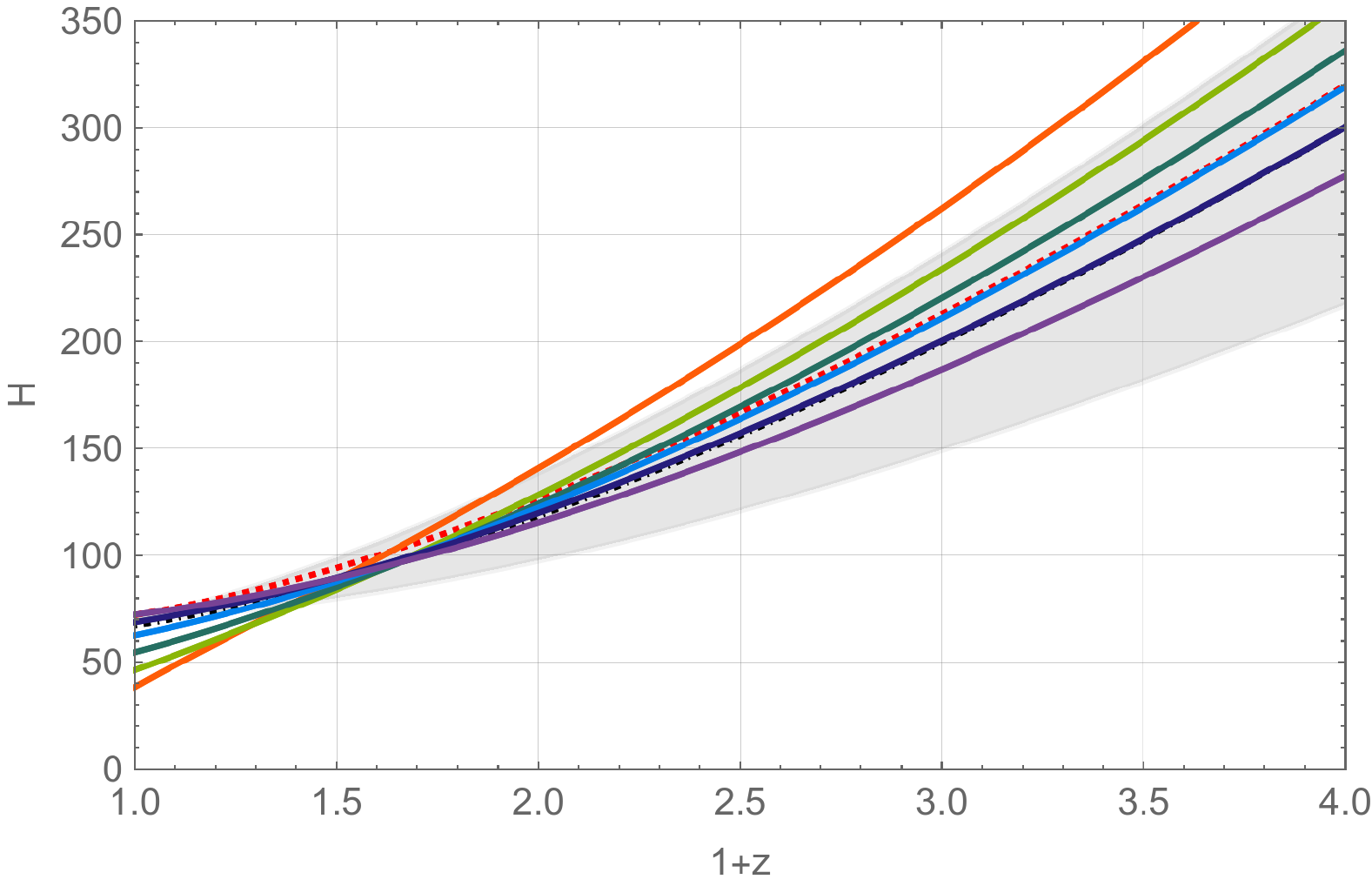}
  \includegraphics[width=.49\linewidth]{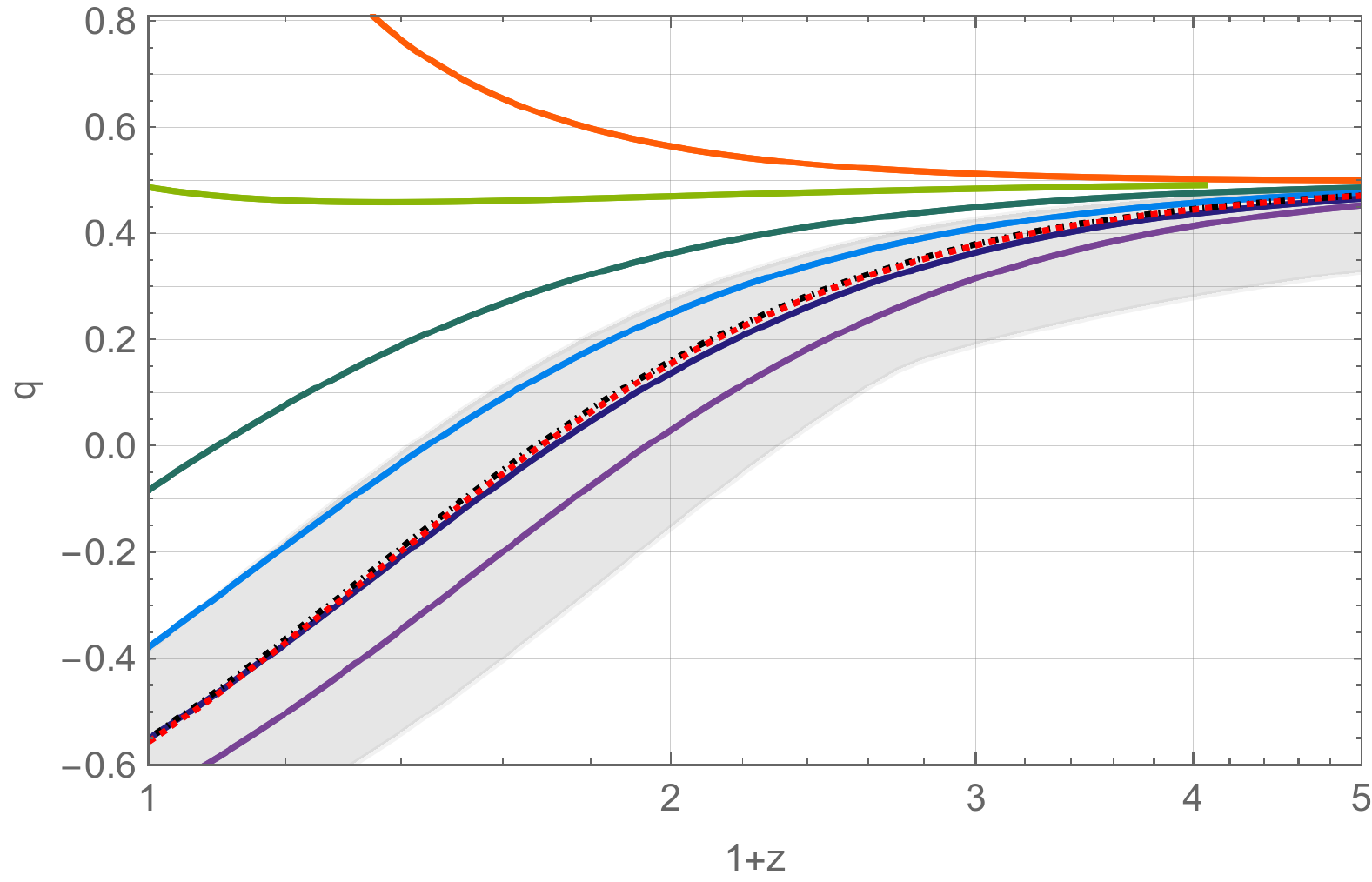}\\
    \includegraphics[width=.49\linewidth]{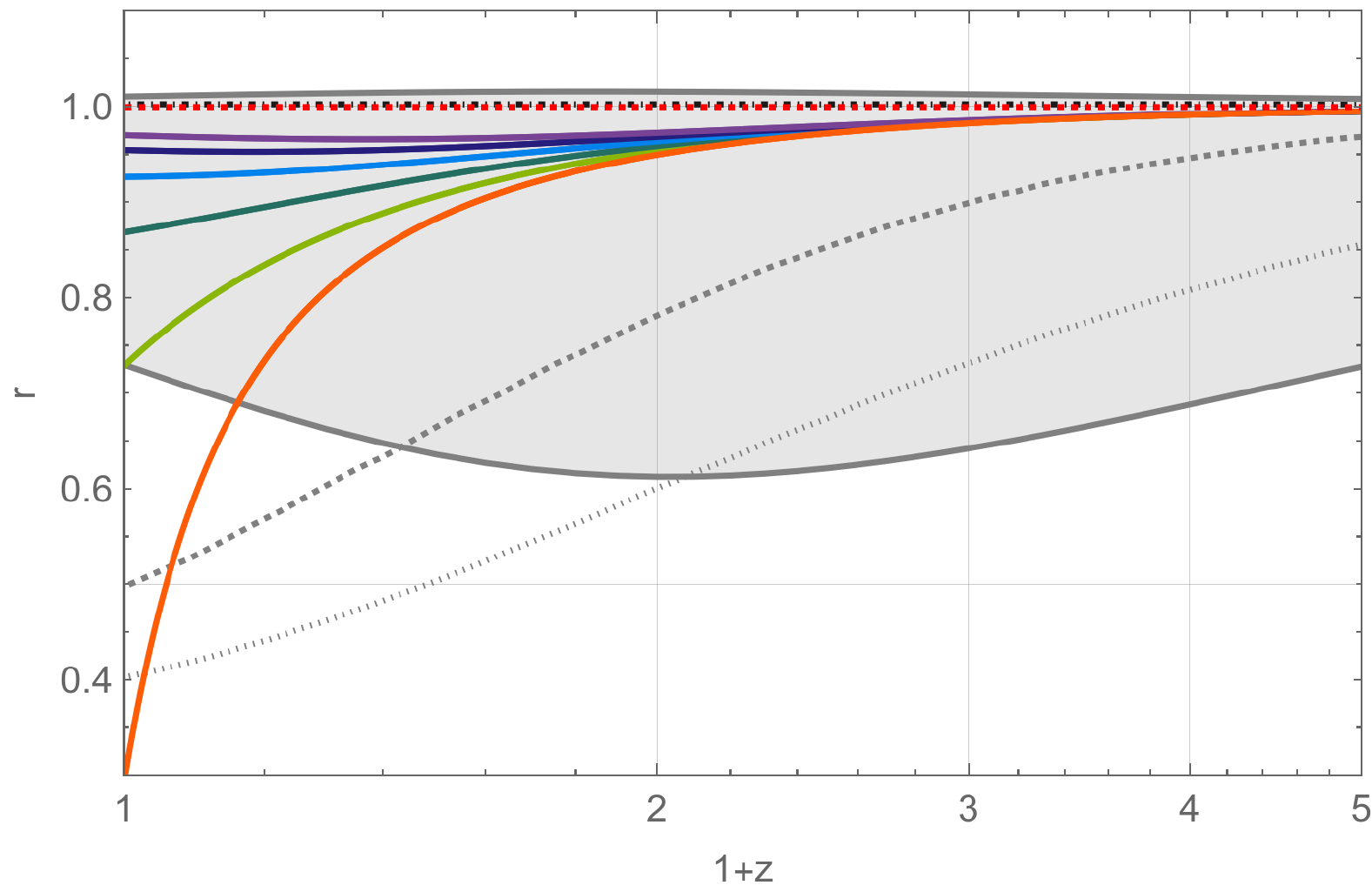}
  \includegraphics[width=.49\linewidth]{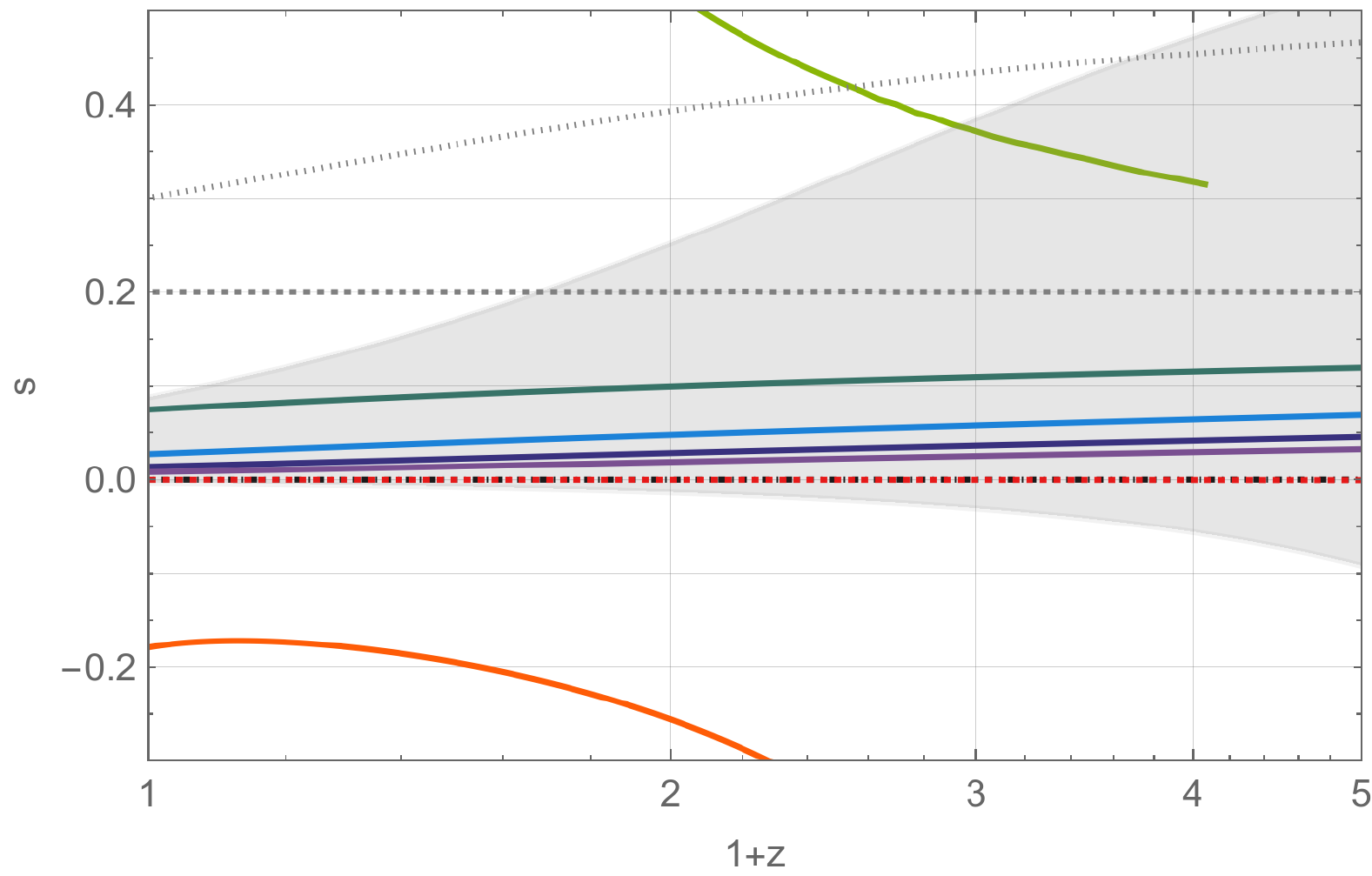}
  \includegraphics[trim={0 4.9cm 0 0},clip,width=.3\linewidth]{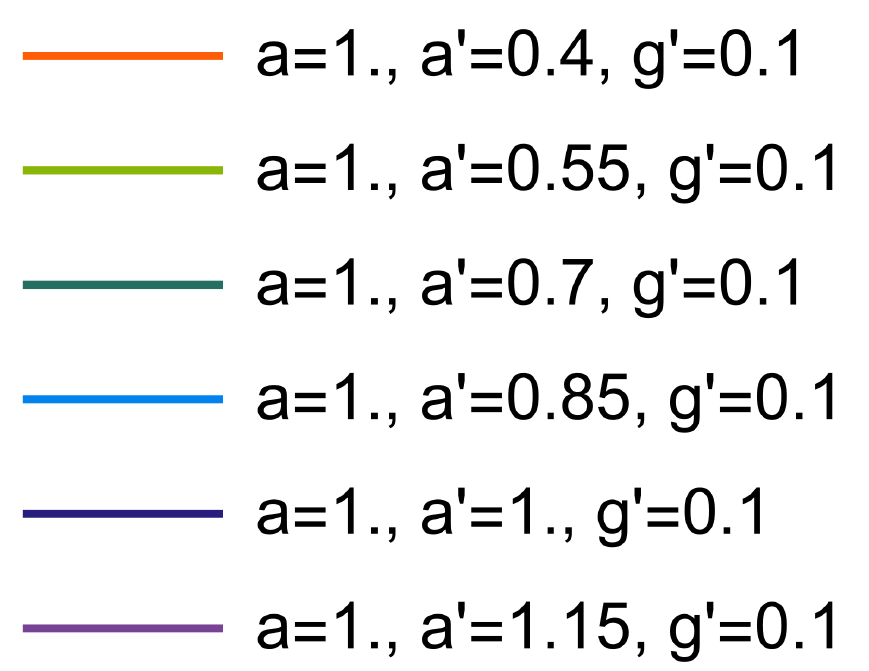}
  \includegraphics[trim={0 0 0 7cm},clip,width=.3\linewidth]{./plot_vary-ap_legend-2.pdf}
\caption{Comparison of the state finder parameters and $H(z)$ for several models with fixed $g'_0 > 0$. The plots are the following: top-left $H(z)$, top-right $q(z)$, bottom-left $r(z)$ and bottom-right $s(z)$. Colored lines correspond to different parameters of the scale dependent cosmology. Gray regions correspond to the allowed parameters in 'Model B' from $H(z)$ and $\mu(z)$ observations. Dashed lines, stating other reference models, are indicated in the legend.}
\label{varyapcase}
\end{figure}

In this section we will comment on the results for a given $g'_0$ positive and several values of $x'_0$. We have considered $g'_0 = 1.0$ and $x'_0 = 0.40, 0.55, 0.70, 0.85, 1.00, 1.15$. Such models have curves for $q$, $r$, $s$ that are easily distinguished in the plots of figure \ref{varyapcase} and do not depend on $h$. The Hubble parameter, however, is not dimensionless and therefore also depends on the scale $h$ of the time variable, see (\ref{tau}). For this reason, the models displayed in the $H(z)$ plot correspond to a value of $h$ such that $t_\text{age}$ determined from (\ref{tage}), give us roughly $t_\text{age} \approx 13.8$ Gyrs See table \ref{tablegp=0.1}.

The transition from deceleration to acceleration occurs at redshift $z_t$ and the present deceleration value is $q_0$, see right-most columns of table \ref{tablegp=0.1}. When $g'_0 \sim 0.1$, all the models with $x'_0 \gtrsim 0.7$ exhibit acceleration/deceleration transition. The models with $x'_0 = 0.70, 0.85, 1.00, 1.15$ have $(z_t, q_0) = (0.10, -0.08)$, $(0.44, -0.38)$, $(0.71, -0.55)$, $(0.94, -0.66)$ respectively. The transition for $\Lambda$CDM occurs at $z_t \approx 0.68$ (Planck values), or using the approximation $a=(\Omega_m/2\Omega_\Lambda)^{1/3}$, at $z_t \approx 0.66 $ using Planck values \cite{Planck:2018vyg}. This approximation uses the model that neglects $\Omega_{rad}$ so that we have the analytic solution 
\begin{equation}
    a(t) = (\Omega_m / \Omega_\Lambda) \sinh ^{2/3}(t/t_\Lambda)\,, \quad t_\Lambda =2/(3 H_0 \sqrt{\Omega_\Lambda}) \,,
\end{equation}
which is fairly accurate for $a>0.01$ (or $t>10$Myrs). The decceleration parameter is given by $q_0 =1/2 \Omega_m-\Omega_\Lambda$ in the $\Lambda$CDM model, and evaluates to $q_0 \approx -0.55$ for the Planck inferred parameters and $q_0 = -0.6 \pm 0.2$ for supernova constraints. For Brans-Dicke model it is obtained $q_0 \approx -0.56$. In \cite{Rahman:2021hrl}, using combined constraints from H(z)+Union 2.1+NVSS-ISW (NRAO VLA Sky Survey and integrated Sachs-Wolfe effect) data the authors obtain $q_0 = -0.5808 \substack{+0.17 \\ -0.13}$ and $z_t = 0.724\pm 0.047$ for $\Lambda$CDM model.

In \cite{Giostri:2012ek}, the authors considered a kink-like parametrization for $q(z)$,\footnote{There are several other well-known parameterizations for the deceleration parameter, see \cite{Mamon:2017rri} and references therein.}
\begin{equation}
    q(z) = q_f + \frac{q_i-q_f}{1-(q_i/q_f)\left(\frac{1+z_t}{1+z}\right)^{1/\tau}}\,.
\end{equation}
The dependence of $r(z)$ deduced from the kink-like $q(z)$ model \footnote{the statefinder $r(z)$ is the same as the cosmographic jerk parameter $j(z)$} is contained in the yellow regions of figures \ref{modelAfig} and \ref{modelBfig}. In \cite{Giostri:2012ek}, the authors used observations from several supernovea datasets, BAO-CMB and CMB to obtain constraints on the parameters of the kink-like model. Their results using combined  SALT2+BAO-CMB observations give $z_t = 0.64 \pm 0.025$ and $q_0 = -0.53 \substack{+0.17 \\ -0.13}$ (here we quoted the 1$\sigma$ confidence intervals). For $x'_0 \gtrsim 0.7$, our model with $g'_0 =0.1$ is compatible with a kink-like shaped $q(z)$. Differences appear in the higher order parameters.

The $r(z)$ parameter is harder to constrain. Values of $x'_0 \gtrsim 0.55$ are fully contained within the constrained parameters of 'Model B', see curves $r(z)$ contained in bottom-left plot of figure \ref{varyapcase}. These curves are also compatible with constraints on parametric reconstructions \cite{AlMamon:2018uby} and non-parametric reconstructions \cite{Mukherjee:2021ggf,Rahman:2021hrl} of the jerk parameter.

The $s(z)$ parameter is depicted in bottom-right plot of figure \ref{varyapcase}. In the plot can be seen that models with $x'_0 \lesssim 0.7$ produce a divergent $s$ for some $z < 5$. This is a reflection of particularity of the scale dependent cosmology (recall that we are considering models were $\Omega_\text{rad} = 0$ exactly) and here the comments after (\ref{spcomment}) apply. Observables that are sensitive on $s(z)$ can be used to constraint the scale dependent cosmology. Models with $x'_0 \gtrsim 0.7$ are reasonably safe within the constrained region of 'Model B'.

\begin{table}[h!]
\centering
\begin{tabular}{c c c l l} 
\multicolumn{1}{c}{} & $x'_0$ & $h$ & $z_t$ & $q_0$ \\ \hline
\multirow{6}{*}{\begin{minipage}{20mm} ~\\ $g_0=1.0\,,$ \\$x_0=1.0\,,$ \\  $g'_0=0.1$ \\ \end{minipage}}  & 0.40   & 0.96 & -- & 1.81 \\ \cline{2-5}
                      & 0.55   & 0.85 & -- & 0.49 \\ \cline{2-5}
                      & 0.70   & 0.79 & 0.10 & -0.08 \\ \cline{2-5}
                      & 0.85   & 0.74 & 0.44 & -0.38 \\ \cline{2-5}
                      & 1.00   & 0.69 & 0.71 & -0.55 \\ \cline{2-5}
                      & 1.15   & 0.63 & 0.94 & -0.66 \\ \hline
\end{tabular}
\caption{Models with $g'_0 > 0$.}\label{tablegp=0.1}
\end{table}

\newpage
\subsection{Case $g'_0<0$}

\begin{figure}[ht]
  \centering
  \includegraphics[width=.49\linewidth]{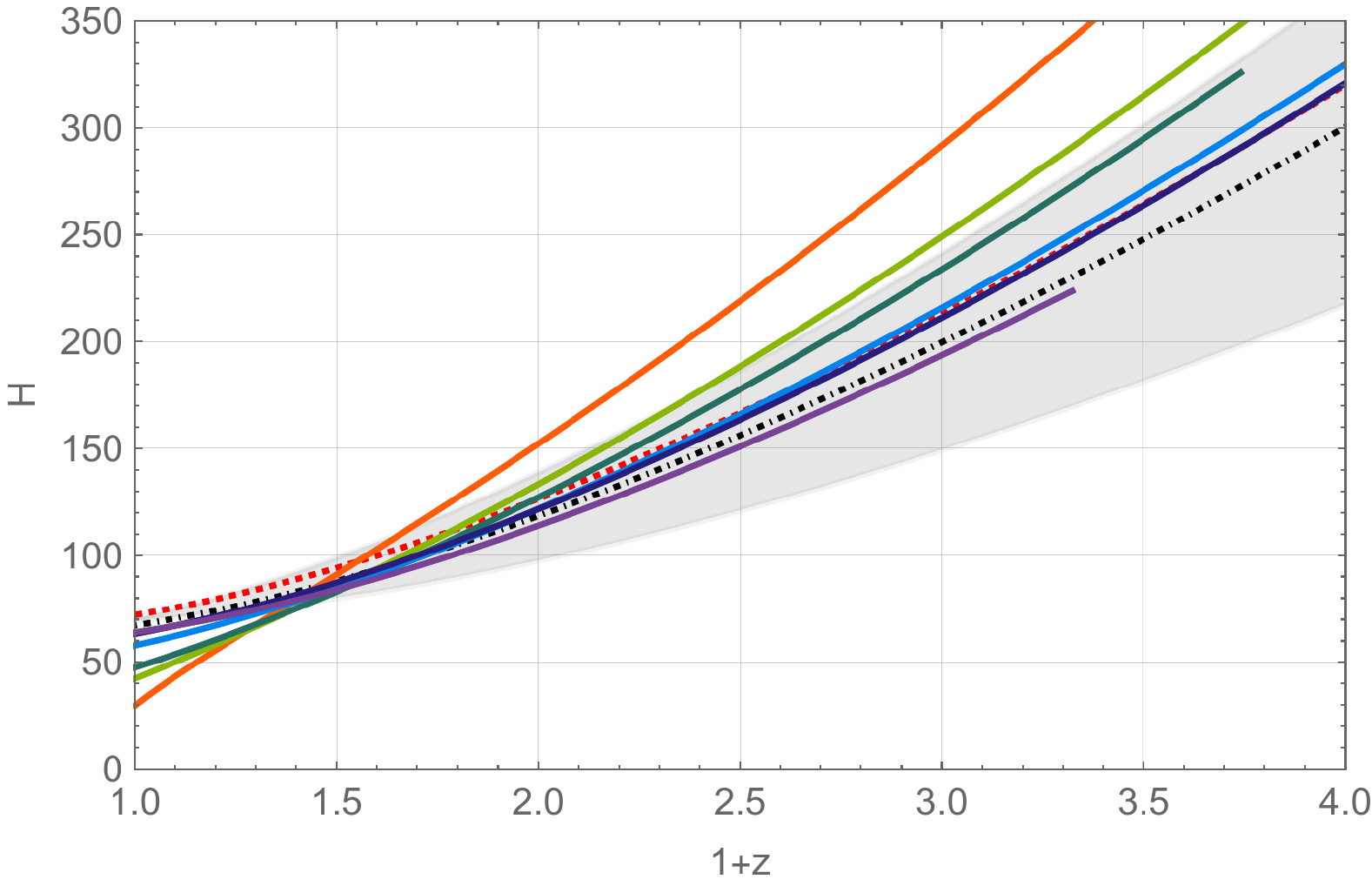}
  \includegraphics[width=.49\linewidth]{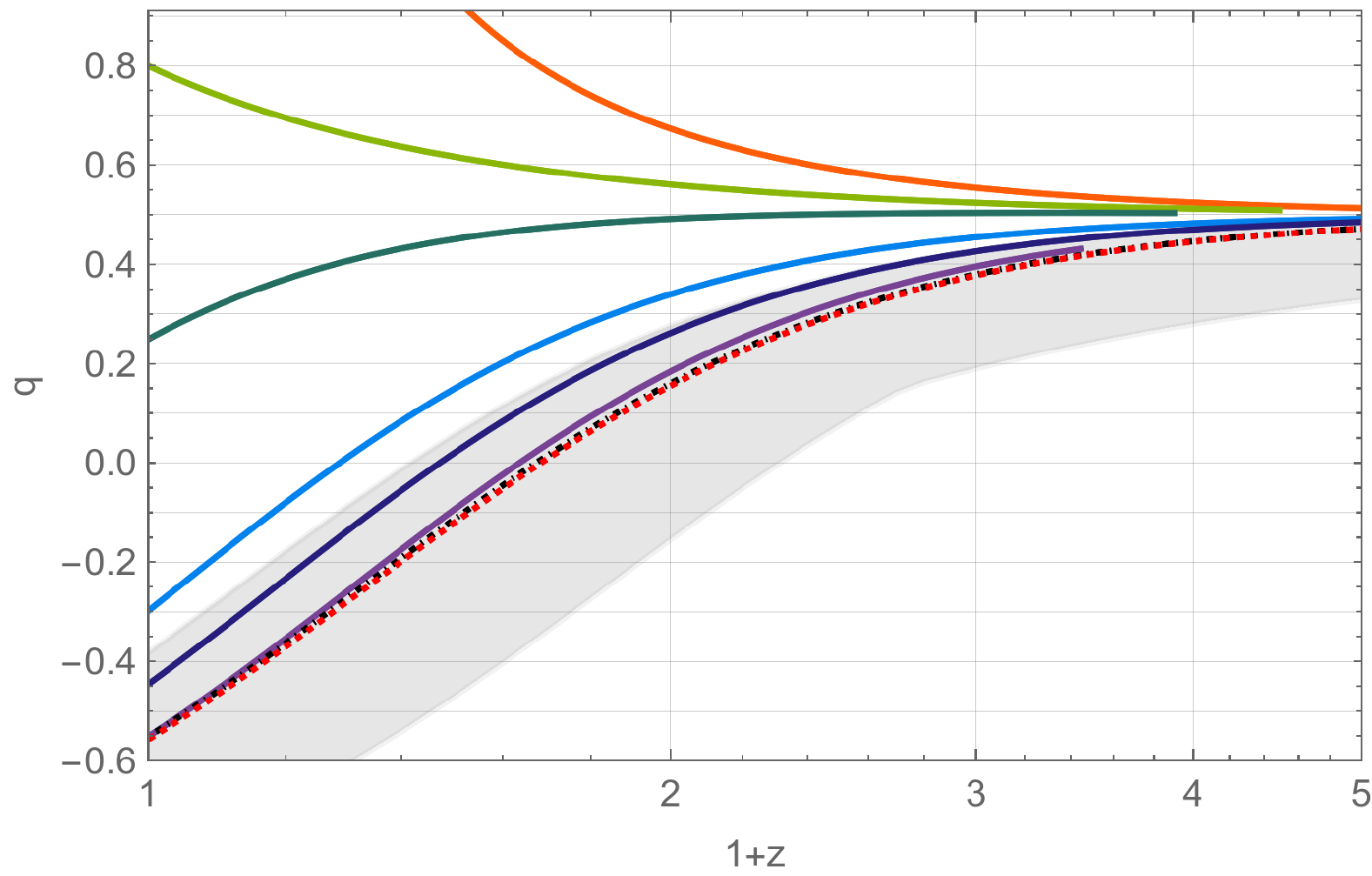}\\
    \includegraphics[width=.49\linewidth]{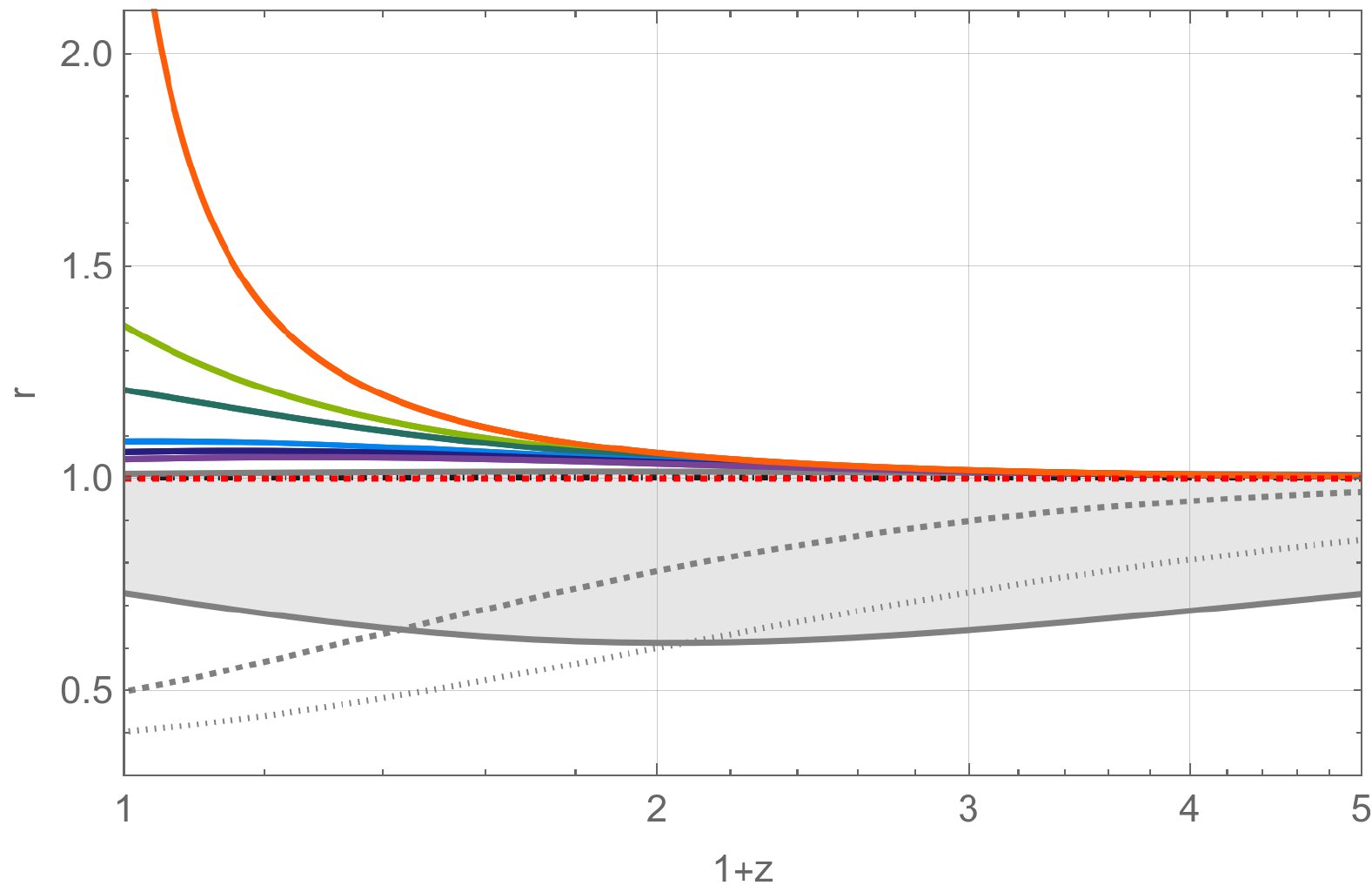}
  \includegraphics[width=.49\linewidth]{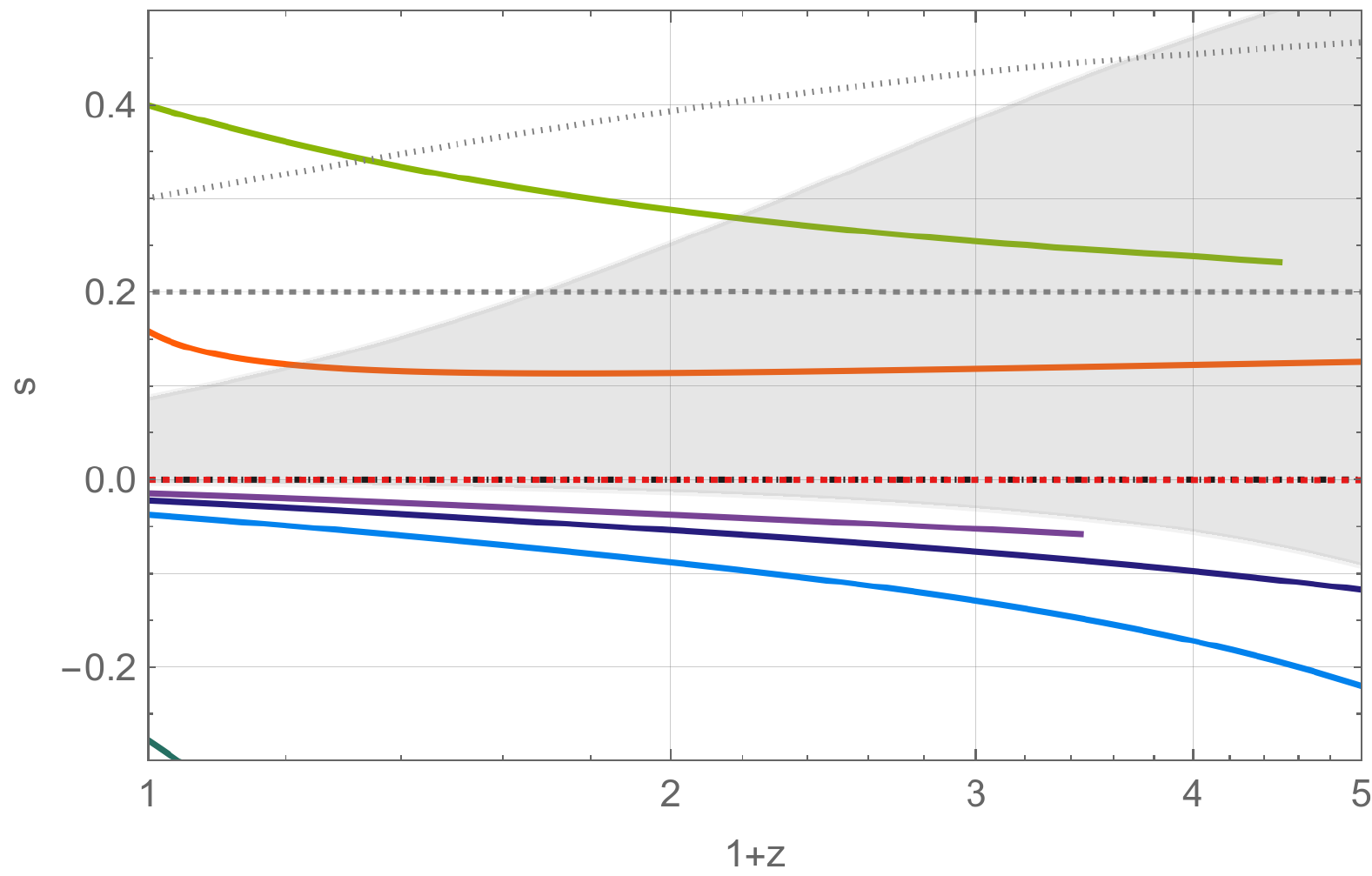}
  \includegraphics[trim={0 4.9cm 0 0},clip,width=.3\linewidth]{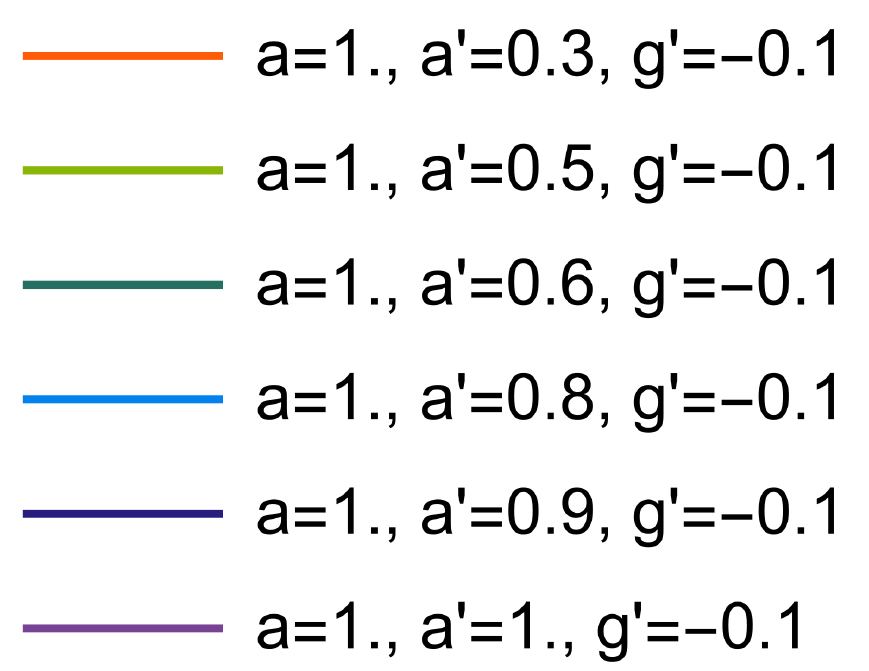}
  \includegraphics[trim={0 0 0 7cm},clip,width=.3\linewidth]{./plot_vary-ap_gpneg_legend-2.pdf}\caption{Comparison of the state finder parameters and $H(z)$ for several models with fixed $g'_0 < 0$.}
\label{varyapcase_gpneg}
\end{figure}

In this section we will comment on the results for a given $g'_0$ negative and several values of $x'_0$. For illustration purposes, we considered $g'_0 = -0.1$. Values of $x'_0 \sim 1$ get close to $\Lambda$CDM in $H(z)$, $q(z)$ and $r(z)$, although $z_t$ and $q_0$ get modified with respect to $\Lambda$CDM. From the bottom-left plot in figure \ref{varyapcase_gpneg}, we can see that the model with negative $g'_0$ produces $r > 1$. From the bottom-right plot we can see that the parameter $s$ exhibits large variations depending on the model. Therefore we can expect that constraints from $\mu(z)$ observations can impose strong restrictions on negative values of $g'_0$ if they are too high. As in the case $g'_0 >0$, here we also see that $z_t$ and $q_0$ depend on the value of $x'_0$, see table \ref{tablegp=-0.1}.

\begin{table}[h!]
\centering
\begin{tabular}{c c c l l} 
\multicolumn{1}{c}{} & $x'_0$ & $h$ & $z_t$ & $q_0$  \\ \hline
\multirow{6}{*}{\begin{minipage}{20mm} ~\\ $g_0=1.0\,,$ \\$x_0=1.0\,,$ \\  $g'_0=-0.1$ \\ \end{minipage}} 
                      & 0.3   & 1.00 & -- & 4.00 \\ \cline{2-5}
                      & 0.5   & 0.85 & -- & 0.80 \\ \cline{2-5}
                      & 0.6   & 0.80 & -- & 0.25 \\ \cline{2-5}
                      & 0.8   & 0.73 & 0.29 & -0.30 \\ \cline{2-5}
                      & 0.9   & 0.71 & 0.47 & -0.44 \\ \cline{2-5}
                      & 1.0   & 0.64 & 0.64 & -0.55 \\ \hline
\end{tabular}
\caption{Models with $g'_0 < 0$.}\label{tablegp=-0.1}
\end{table}

\newpage
\subsection{Case $a'_0>0$}

\begin{figure}[ht]
  \centering
  \includegraphics[width=.49\linewidth]{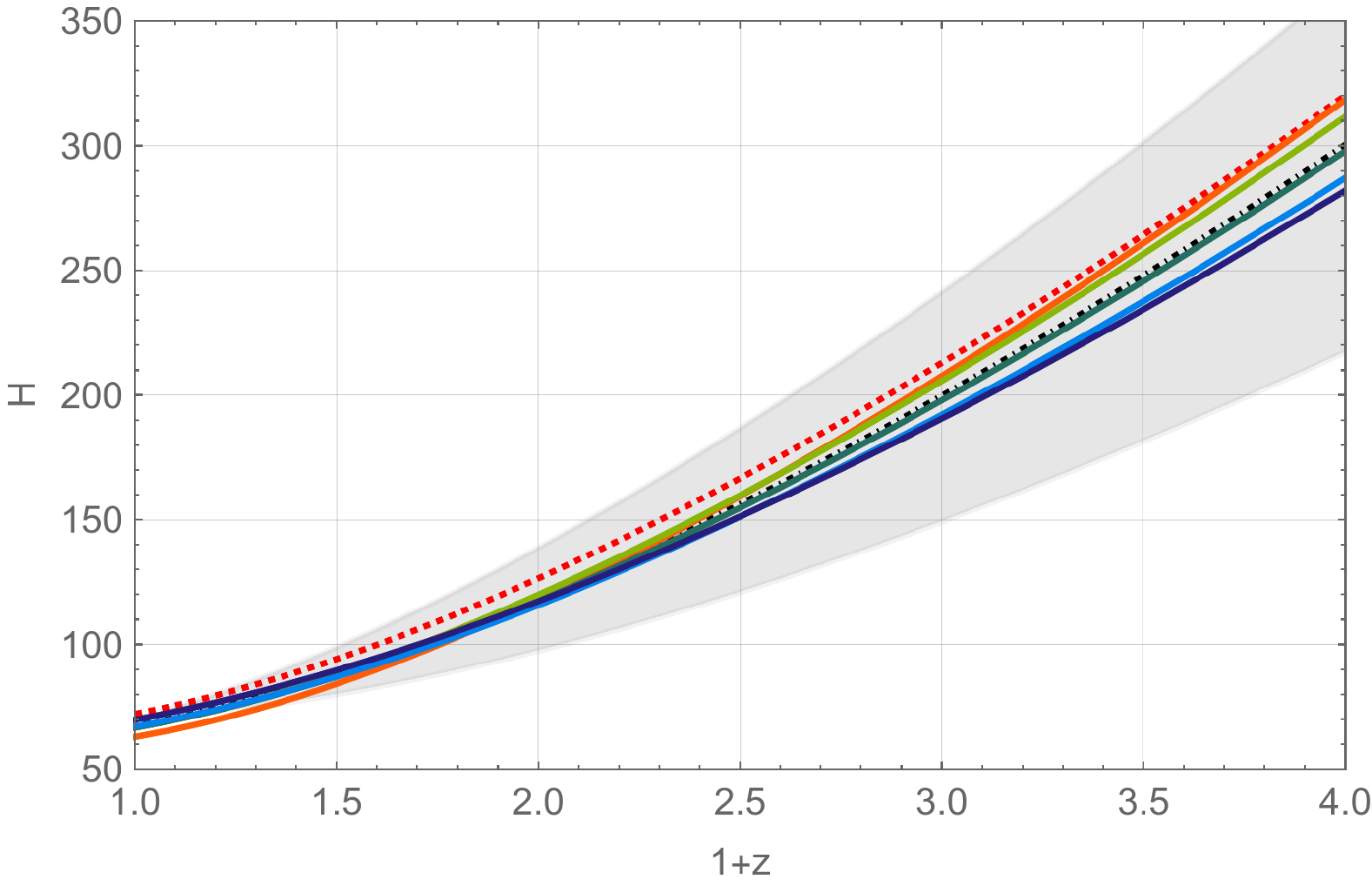}
  \includegraphics[width=.49\linewidth]{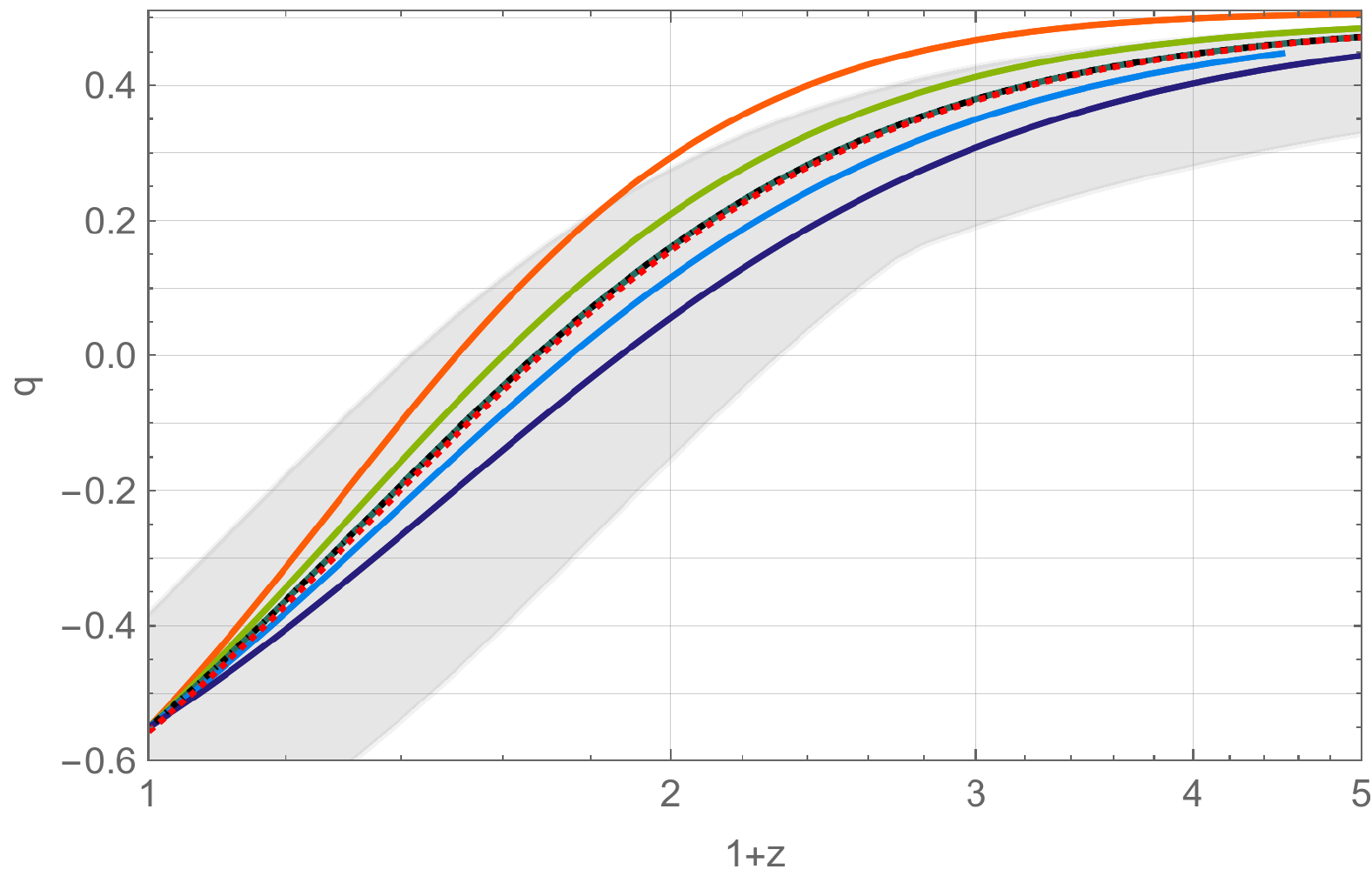}\\
    \includegraphics[width=.49\linewidth]{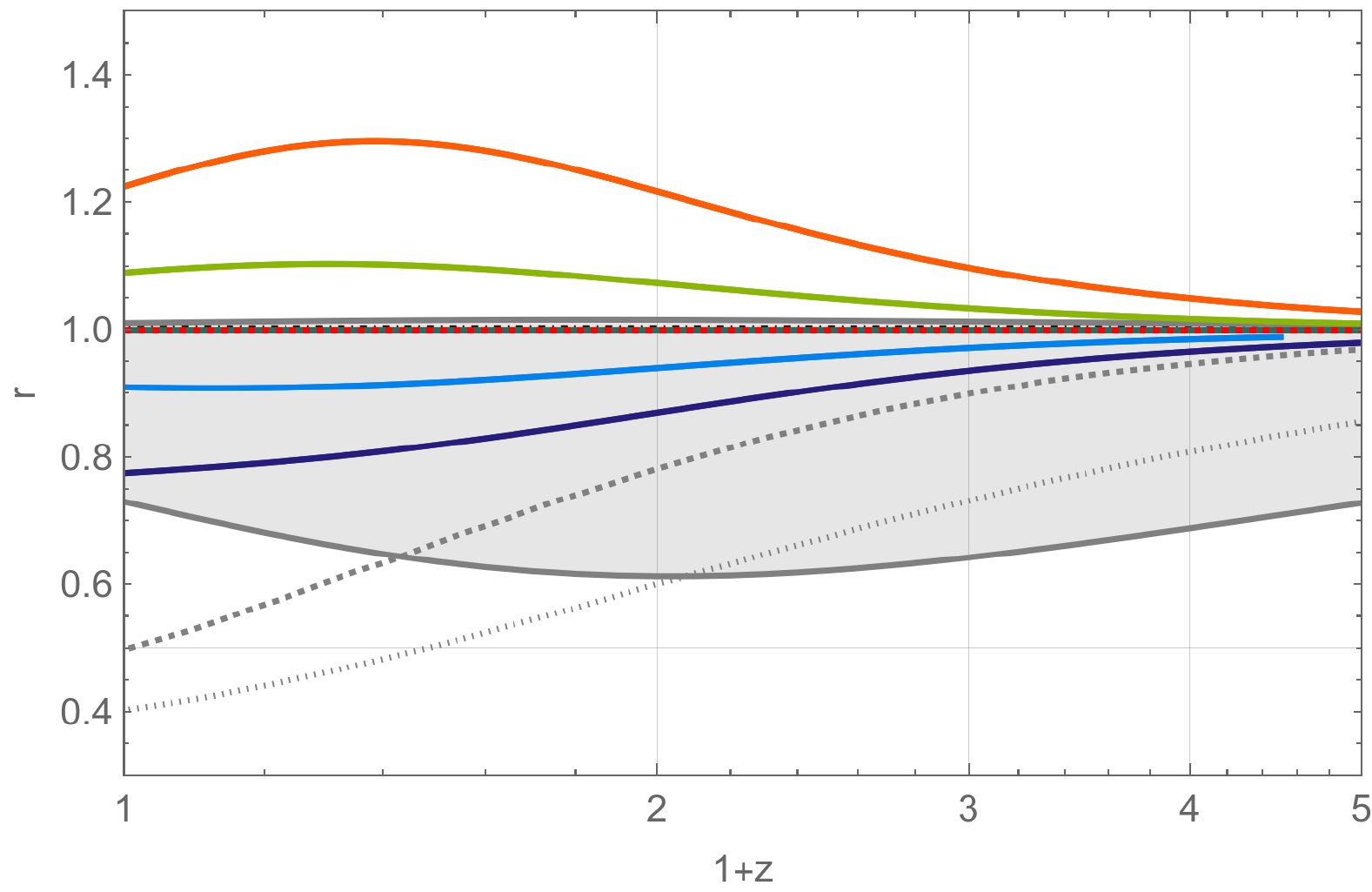}
  \includegraphics[width=.49\linewidth]{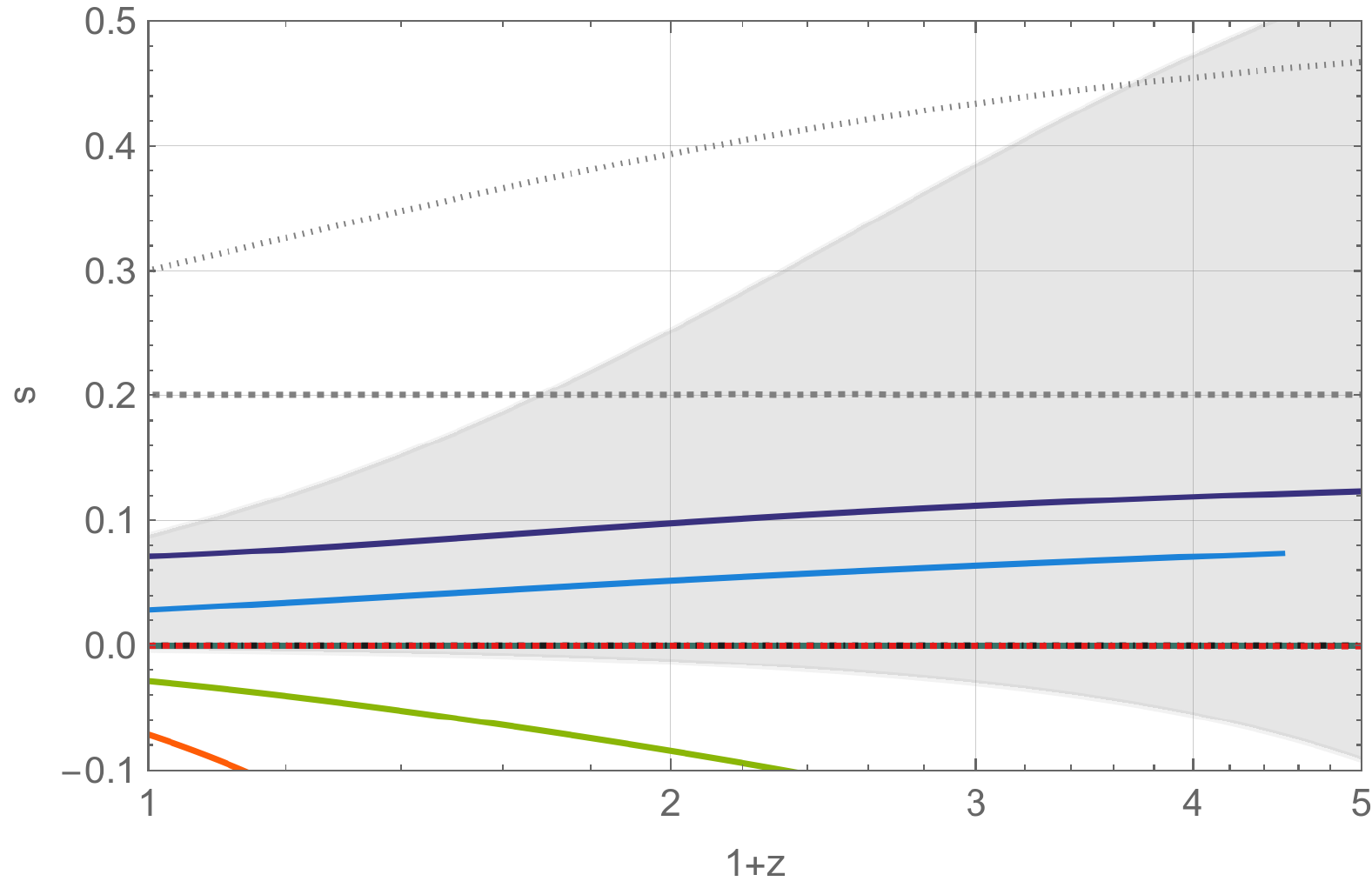}
  \includegraphics[trim={0 4.9cm 0 0},clip,width=.3\linewidth]{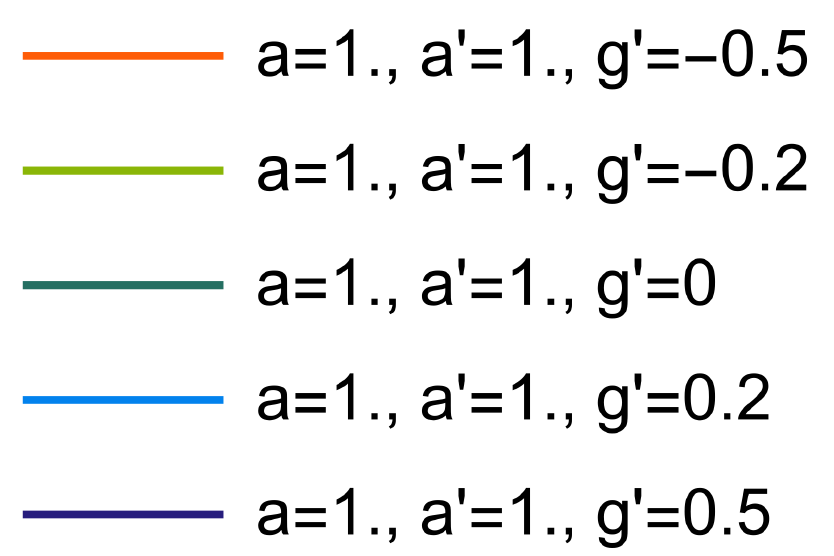}
  \includegraphics[trim={0 0 0 7cm},clip,width=.3\linewidth]{./plot_vary-gp_legend-2.pdf}
\caption{Comparison of the state finder parameters and $H(z)$ for several models with fixed $x'_0 = 1$.}
\label{varygpcase}
\end{figure}

In this section, we will comment on the results for a given $a'_0$ positive and several values of $g'_0$. We have consider models where $g'_0 \approx -0.5, -0.2, 0, 0.2, 0.5$. Plots of figure \ref{varygpcase} make it clear that the scale-dependent model can get arbitrarily close to $\Lambda$CDM for $|g'_0|$ sufficiently small.

Plot $q(z)$ of figure \ref{varygpcase} shows that models where $x'_0$ is chosen appropriately can be such that $q_0$ and $q_{z\rightarrow \infty}$ match the $\Lambda$CDM prediction. The redshift acceleration/deceleration transition gets modified, however, $z_t > z_t^{(\LCDM)}$ for $g'_0 > 0$ and $z_t < z_t^{(\LCDM)}$ for $g'_0 < 0$, see table \ref{tablefixedxp}.

Plots $r(z)$ and $s(z)$ of figure \ref{varygpcase} can be used as an indication that $\mu(z)$ observations will prefer positive values of $g'_0$.

\begin{table}[h!]
\centering
\begin{tabular}{c c c l l} 
\multicolumn{1}{c}{} & $g'_0$ & $h$ & $z_t$ & $q_0$ \\ \hline
\multirow{6}{*}{\begin{minipage}{20mm} ~\\ $g_0=1.0\,,$ \\$x_0=1.0\,,$ \\  $x'_0=1.0$ \\ \end{minipage}} 
                      & -0.5   & 0.63 & 0.51 & -0.55 \\ \cline{2-5}
                      & -0.2   & 0.67 & 0.59 & -0.55 \\ \cline{2-5}
                      & 0   & 0.67 & 0.66 & -0.55 \\ \cline{2-5}
                      & 0.2   & 0.67 & 0.76 & -0.55 \\ \cline{2-5}
                      & 0.5   & 0.70 & 0.86 & -0.55 \\ \hline
\end{tabular}
\caption{Models with $x'_0 =1$.}\label{tablefixedxp}
\end{table}

\section{Conclusions and final remarks}

In this work, we reported statefinder parameters as functions of redshift for the scale-dependent cosmological model proposed in \cite{Alvarez:2020xmk}. We provided analytical expressions that are helpful in finding numerical solutions. Our results provide a good guide for finding appropriate priors for the determination of maximum likelihood contours. Our study shows that for certain regions of the parameter space, the SD model predicts $H(z)$ and $q(z)$ very close to Brans-Dicke or $\Lambda$CDM models. Deviations can be seen at the level of $r(z)$, but this parameter has weaker observational constraints.


The deceleration $q(z)$ and jerk parameter $r(z)$ allowed us to see an important property of the SD model: for generic initial conditions, the SD model has convergence to the $\Lambda$CDM model for high redshift. Furthermore, the $r$-curves explored in  figures~\ref{varyapcase} and~\ref{varyapcase_gpneg} allowed us to see that upper and lower limits for $g'_0>0$ and $g'_0<0$ can be deduced from $\mu(z)$ observations. We propose such a study for a future work.


Our results for the jerk parameter $r$ indicate (see e.g. bottom-left panels in figures~\ref{varyapcase} and~\ref{varyapcase_gpneg}) that there exist cases where $r(z)$ will remain very close to $\Lambda$CDM  for sufficiently big $a'_0$. Also, for the conditions given in figure~\ref{varygpcase}, it can be seen that $r(z)$ peaks at low redshifts $z \lesssim 1$, which is a particular behavior present in the SD model and is not observed in the other theories considered in this work.

The comparison with the allowed parameters in 'Model B' inferred from the distance modulus $\mu(z)$ observations in the statefinders diagnostic suggests the constraints $a' \gtrsim 0.6$ if $g'_0 \sim  0.1$ and $a' \gtrsim 0.9$ if $g'_0 \sim  -0.1$. This set of constraints is thus helpful when defining realistic initial conditions of the SD scenario of gravity.

Additionally, we determine deceleration-acceleration transition $z_t$ for several initial conditions. 
For models such that $q_0 \approx q_0(\LCDM) \in (-0.55, -0.6)$, the transition $z_t$ is in the range $z_t \sim (0.7,0.9)$ if $g'_0 \sim 0.1$ or $z_t \sim (0.64, 0.7)$ if $g'_0 \sim -0.1$. Smaller $|g'_0|$ values would give smaller deviations with respect to $z_t(\LCDM)$. It is interesting to note that the viscous fluid models also present a slightly earlier decelerated-accelerated transition compared with the one observed in the standard $\Lambda$CDM model, with values $z_t\approx 0.755$~\cite{Herrera-Zamorano:2020rdh}. Let us remark that the SD model does not include an early accelerated phase (besides the accelerated phase that begins at $z_t$) at higher redshifts such as the one at $z \sim 17.2$, as it has been observed in some Einstein-Gauss-Bonnet~\cite{Garcia-Aspeitia:2020uwq} and $f(R)$ models~\cite{Jaime:2018ftn}




The present study can be extended in various directions:
\newline

\textit{Observational constraints} provided by $H(z)$, $\mu(z)$ and BAO-CMB observations can be used to obtain maximum likelihood contours for the parameters of the SD model. The information provided in this paper is useful, and we are currently working on that topic for a separate publication.  
\newline

\textit{SD scenario from a variational principle.} The present model uses a NEC to close the system in~(\ref{SD1})-(\ref{NECCosm}). 
From a theoretical point of view, it would
be interesting to explore the explicit form
the possible actions that give to 
these equations.
Attempts to derive this condition from an action require a non-trivial constraint on the Einstein-frame Ricci tensor from worldsheet string theory, see for instance, \cite{Parikh:2014mja}. 
If we want to extend the current approach to field theories that do not satisfy some well-known energy conditions (including the NEC), we must look for an alternative route to close the system. One possible pathway relies upon fixing the renormalization scale in terms of the field variables by a variational procedure and then putting the scale on-shell~\cite{Koch:2010nn,Koch:2014joa,Koch:2020baj}.
Inputs for the beta functions of the couplings coming from quantum gravity could allow generating an extra gap equation, changing the dynamics of the SD scenario of cosmology. The inclusion of these quantum effects will be left for future work.
\newline

\textit{SD Brans-Dicke and K-essence and their relation with positivity bounds}. By promoting the couplings in the BD and K-essence approaches to scale-dependent quantities, one can derive cosmological models that realize Weinberg's asymptotic safety scenario, see for instance \cite{Reuter:2003ca,Wetterich:2014gaa}. It is interesting to study cosmological constraints of the effective parameters in those models in a reverse version of the positivity bounds program \cite{Adams:2006sv,deRham:2017avq,deRham:2017zjm}. The latter uses unitarity, causality, and locality of effective field theories (EFT) to derive bounds on the EFT parameters that translate into stronger constraints parameters of the model \cite{Melville:2019wyy}.

\section*{Acknowlegements}

C.L. acknowledges the scholarship Becas Chile, ANID-PCHA/2020-72210073 for financial support. The author A. R. acknowledges Universidad de Tarapacá for financial support. 

\bibliography{References.bib}

\end{document}